\documentclass[a4paper,11pt]{article}
\pdfoutput=1 % if your are submitting a pdflatex (i.e. if you have
\usepackage{jheppub} % for details on the use of the package, please
\usepackage[T1]{fontenc} % if needed
\usepackage{mathrsfs,amsthm,enumerate,float,ytableau}
\usepackage[utf8]{inputenc}
\usepackage{lmodern}
\newcommand{\al}{\alpha}
\newcommand{\be}{\beta}
\newcommand{\de}{\delta}

\newcommand{\vep}{\varepsilon}

\newcommand{\ka}{\kappa}
\newcommand{\la}{\lambda}
\newcommand{\om}{\omega}
\newcommand{\si}{\sigma}
\renewcommand{\th}{\theta}
\newcommand{\vp}{\varphi}

\newcommand{\ze}{\zeta}
\newcommand{\De}{\Delta}

\newcommand{\La}{\Lambda}
\newcommand{\Si}{\Sigma}

\newcommand{\bde}{\boldsymbol{\delta}}
\newcommand{\bev}{\mathbf e}

\newcommand{\bk}{\mathbf{k}}

\newcommand{\bn}{\mathbf{n}}

\newcommand{\bbs}{\mathbf{s}}

\newcommand{\bx}{\mathbf{x}}

\newcommand{\bsi}{{\boldsymbol{\si}}}
\newcommand{\bz}{\mathbf{z}}
\newcommand{\bxi}{\boldsymbol{\xi}}

\newcommand{\bnu}{\boldsymbol{\nu}}
\newcommand{\bell}{\boldsymbol\ell}
\newcommand{\tK}{\widetilde{K}}

\newcommand{\tS}{\widetilde{S}}

\newcommand{\tf}{\widetilde{f}}

\newcommand{\wth}{\widetilde{h}}

\newcommand{\tn}{\widetilde{n}}

\newcommand{\tnu}{\widetilde\nu}

\newcommand{\CC}{{\mathbb C}}
\newcommand{\NN}{{\mathbb N}}
\newcommand{\RR}{{\mathbb R}}
\newcommand{\ZZ}{{\mathbb Z}}
\newcommand{\cB}{{\mathcal B}}

\newcommand{\cE}{{\mathcal E}}

\newcommand{\cH}{{\mathcal H}}

\newcommand{\cP}{{\mathcal P}}
\newcommand{\cR}{{\mathcal R}}

\newcommand{\cT}{{\mathcal T}}

\newcommand{\cZ}{{\mathcal Z}}
\newcommand{\fH}{{\mathfrak H}}

\newcommand{\oH}{\overline\cH}
\newcommand\Hsc{H_{\mathrm{sc}}}
\newcommand\Hsp{H_{\mathrm{spin}}}
\newcommand\Zsc{Z_{\mathrm{sc}}}

\newcommand{\pd}{\partial}

\newcommand{\ket}[1]{|#1\rangle}

\let\ds\displaystyle

\newcommand{\mss}{\kern 1pt}

\renewcommand{\le}{\leqslant}
\renewcommand{\ge}{\geqslant}
\newcommand{\tends}[1]{\bbuildrel{\hbox to 2em{\rightarrowfill}}_{#1}^{}}

\newcommand{\mP}{\mathscr P}
\newcommand{\erf}{\operatorname{erf}}
\newcommand{\sgn}{\operatorname{sgn}}

\newcommand{\tr}{\operatorname{tr}}

\newcommand{\iu}{\mathrm i}
\newcommand{\e}{\mathrm e}

\newcommand{\su}{\mathrm{su}}

\newcommand{\gl}{\mathrm{gl}}
\renewcommand{\Im}{\operatorname{Im}}
\newcommand{\sn}{\operatorname{sn}}
\newcommand{\cn}{\operatorname{cn}}

\newcommand{\dn}{\operatorname{dn}}

\newcommand{\en}{\enspace}
\newcommand{\all}{\forall}

\newcommand{\Int}[1]{\,\mathop{\!#1}\limits^{\lower1ex\hbox{$\scriptstyle\circ$}}{}}

\newcommand{\ms}{\mskip1mu}

\theoremstyle{remark}
\newtheorem{rem}{Remark}

\def\clap#1{\hbox to 0pt{\hss#1\hss}}

\def\mathclap{\mathpalette\mathclapinternal}

\def\mathclapinternal#1#2{%
           \clap{$\mathsurround=0pt#1{#2}$}}

         \title{A novel class of translationally invariant\\ spin chains with long-range
           interactions}

\author[a]{B. Basu-Mallick,}
\author[b]{F. Finkel,}
\author[b,1]{A. Gonz\'alez-L\'opez,\note{Corresponding author.}}

\affiliation[a]{Saha Institute of Nuclear Physics, Theory Division, HBNI, 1/AF Bidhan Nagar,\\
  Kolkata 700 064, INDIA}
\affiliation[b]{Universidad Complutense de Madrid, Depto.~de F\'\i sica Te\'orica,
  Plaza de las Ciencias 1,\\28040 Madrid, SPAIN}

\emailAdd{bireswar.basumallick@saha.ac.in}
\emailAdd{ffinkel@ucm.es}
\emailAdd{artemio@ucm.es}

\abstract{We introduce a new class of open, translationally invariant spin chains with long-range
  interactions depending on both spin permutation and (polarized) spin reversal operators, which
  includes the Haldane--Shastry chain as a particular degenerate case. The new class is
  characterized by the fact that the Hamiltonian is invariant under ``twisted'' translations,
  combining an ordinary translation with a spin flip at one end of the chain. It includes a
  remarkable model with elliptic spin-spin interactions, smoothly interpolating between the XXX
  Heisenberg model with anti-periodic boundary conditions and a new open chain with sites
  uniformly spaced on a half-circle and interactions inversely proportional to the square of the
  distance between the spins. We are able to compute in closed form the partition function of the
  latter chain, thereby obtaining a complete description of its spectrum in terms of a pair of
  independent $\su(1|1)$ and $\su(m/2)$ motifs when the number $m$ of internal degrees of freedom
  is even. This implies that the even $m$ model is invariant under the direct sum of the Yangians
  $Y(\gl(1|1))$ and $Y(\gl(0|m/2))$. We also analyze several statistical properties of the new
  chain's spectrum. In particular, we show that it is highly degenerate, which strongly suggests
  the existence of an underlying (twisted) Yangian symmetry also for odd $m$.}

\begin{document}

\keywords{Lattice Integrable Models [100], Quantum Groups [20]}

\maketitle
\flushbottom

\section{Introduction}

Exactly solvable quantum dynamical models in one dimension and spin chains with long-range
interactions have given rise to interesting applications as prototypes of strongly correlated
systems exhibiting generalized exclusion statistics~\cite{Ha91b,HHTBP92,GS05,Gr09}, as well as in
connection with such diverse phenomena as the quantum Hall effect~\cite{AI94,BK09}, quantum
transport in mesoscopic systems~\cite{BR94,Ca95} or quantum simulation of long-range
magnetism~\cite{HGCK16}. Moreover, recent experiments involving optical lattices of ultracold
Rydberg atoms and trapped ions have opened up the exciting possibility of experimentally probing
various properties of such low-dimensional quantum spin chains in a very precise
way~\cite{PC04,GL14,KCIKD09,RGLSS14,JLHHZ14,SZFHC15}. In the context of high-energy physics,
quantum spin chains with long-range interactions have played a key role in the computation of the
spectrum of the dilation operator in $N=4$ super Yang--Mills theories~\cite{MZ03,SS04,Be12}.
Intriguing connections between (1+1)-dimensional conformal field theory (CFT) and exactly solvable
models with long-range interactions have also been uncovered in recent
years~\cite{CS10,NCS11,BQ14,TNS14}. In the same vein, it has been found that the Casimir equation
for conformal blocks in $d$ dimensions can be transformed into the eigenvalue problem for a
hyperbolic Calogero--Sutherland model associated with the $BC_N$ reflection
group~\cite{IS16,ILLS18}.

The translationally invariant Haldane--Shastry (HS) spin-$\frac{1}{2}$ chain~\cite{Ha88,Sh88},
whose spins are uniformly spaced on a circular lattice and exhibit two-body interactions inversely
proportional to the square of their chord distances, is the prototypical example of an exactly
solvable lattice model with long-range interactions. The HS spin chain features many interesting
properties, a few of which we shall briefly review next. Indeed, its exact ground state coincides
with the $U\to\infty$ limit of Gutzwiller's variational wavefunction for the Hubbard model, and
with the one-dimensional version of Anderson's ``resonating-valence-bond
state''~\cite{An73,FA74,ABZH87,Ha88}. Moreover, its spinon excitations~\cite{Ha91} (in the
antiferromagnetic case) can be described through a generalized Pauli exclusion principle, as they
can be regarded as anyons with exclusion parameter $g=1/2$ (so called semions)~\cite{Ha91b}. The
original ---$\su(2)$--- HS spin chain and its $\su(m)$ generalization also exhibit Yangian quantum
group symmetry, even for a finite number of lattice sites~\cite{HHTBP92,BGHP93}. As a consequence,
the spectrum of these integrable spin chains can be described by means of finite sequences of the
binary digits 0 and 1, known as ``motifs'' in the literature~\cite{HHTBP92}, related to certain
finite-dimensional representations of the Yangian algebra~\cite{KKN97,NT98}. In addition, the
complete energy spectrum of these chains, including the degeneracy of each level, can be generated
from the energy function of an associated one-dimensional classical vertex model~\cite{BBH10}. The
$\su(m)$ HS chain can also be obtained by taking the freezing limit~\cite{Po93} of the $\su(m)$
spin Sutherland model~\cite{HH92}, whose particles possess both dynamical and spin degrees of
freedom. More precisely, in the large coupling constant limit the spin part of the latter model's
Hamiltonian decouples from the dynamical one and yields the Hamiltonian of the HS chain. In fact,
the partition function of the $\su(m)$ HS chain can be computed in closed form~\cite{FG05} by
taking advantage of this decoupling of the spin and dynamical degrees of freedom of the spin
Sutherland model, which is the essence of Polychronakos's freezing trick.

In view of these remarkable properties, several methods have been proposed for constructing
integrable or exactly solvable variants of the HS chain. In particular, the connection between
quantum integrable models with long-range interactions and different (extended) root systems has
proved a useful tool in this endeavor~\cite{OP83,CS02}. The main idea in this respect is to
replace the $A_{N-1}$ root system associated with the interaction potential of the spin Sutherland
model, which yields the original HS chain, by other (extended) root systems like, e.g., $BC_N$,
$B_N$ and $D_N$. The spin Sutherland models associated with all of these root systems have in fact
been studied in the literature, and used to construct spin chains of HS type by taking their
strong coupling limits~\cite{BPS95,EFGR05,BFG11,BFG13}. In particular, the exact partition
function of each of these chains has been computed in closed form applying the freezing trick to
their parent spin dynamical models. It should be noted, however, that none of these integrable
variants of the HS chain retains the translational invariance characteristic of the original
model.

Several extensions of the HS spin chain with lattice sites arbitrarily distributed on a circle and
exact ground state wavefunctions have also been recently constructed using infinite matrix product
states (MPSs)~\cite{CS10} related to certain rational CFTs. More precisely, it has been observed
that the correlators associated with the chiral vertex operators of such CFTs play the role of
infinite MPSs yielding the ground state wavefunctions of these HS-like
chains~\cite{CS10,NCS11,BQ14,TNS14}. Using the null field techniques related to rational CFTs, one
can explicitly construct the Hamiltonians of these models and study interesting physical
properties thereof, like, e.g., the spin-spin correlation functions. However, these models lack
two of the key properties of the original HS chain, namely, they do not exhibit translational
invariance nor are they integrable in general. More recently, exact ground state wavefunctions of
some HS-like spin chains, with open boundary conditions and lattice points arbitrarily distributed
on a half-circle, have been constructed using infinite MPSs related to suitable boundary CFTs and
corresponding null fields~\cite{TS15}. Again, these CFT-inspired generalizations of the HS chain
with open boundary conditions become integrable~\cite{TS15} and, in fact, exactly
solvable~\cite{BFG16}, only for some special choices of the lattice points, but none of them
possess the translational invariance of the original HS chain.

In this work we construct a new solvable spin chain with inverse-square long-range interactions
possessing two of the essential properties of the $\su(m)$ HS chain, namely translational
invariance and uniform spacing of the sites. Our construction is still based on applying
Polychronakos's freezing trick to an appropriate (solvable) translationally invariant spin
dynamical model~\cite{FGGRZ01}. However, unlike the original HS chain and its $BC_N$, $B_N$ and
$D_N$ versions mentioned above, neither this dynamical model nor the resulting chain are
associated with an extended root system in the standard fashion. More precisely, the spin chain we
shall construct is translationally invariant, in the sense that (like the original HS chain of
$A_{N-1}$ type) the interactions between sites $i$ and $j$ depends only on the difference $i-j$.
However, in contrast with the latter chain, its Hamiltonian includes both spin permutation and
spin flip operators generating the full Weyl group of $D_N$ type. In particular, the presence of
the spin flip operators in the Hamiltonian also distinguishes the new model from the spin chains
constructed from infinite MPSs in the above cited references. On the other hand, we shall see that
the new chain's sites are uniformly arranged on a half-circle, like in uniformly spaced HS chains
of $BC_N$ type, so that the model is naturally regarded as an open chain. As such, the usual
translation operator does not commute with the Hamiltonian due to boundary terms. Remarkably,
however, the new model is exactly invariant under a cyclic group of ``twisted'' translations,
suitably combining standard translations with spin flips. This symmetry is indeed one of the
hallmarks of the model, which we believe sets it apart from all the solvable spin chains with
long-range interactions discussed above.

In fact, our construction can be substantially extended in two different directions. In the first
place, we can replace the standard spin flip operators in the Hamiltonian by partially polarized
versions thereof without compromising any of the key properties discussed above, namely
translational invariance of the interactions, uniform spacing of the sites, symmetry under twisted
translations and solvability. Secondly, we can allow for much more general spin-spin interactions
such that the resulting models still possess the characteristic symmetry under twisted
translations discussed above. This yields a wide class of new open, translationally invariant spin
chains with long-range interactions and uniformly spaced sites, whose solvability properties are
certainly worth investigating. This class includes a remarkable model with elliptic interactions,
reminiscent of the well known Inozemtsev~\cite{In90} chain, which smoothly interpolates between
the new solvable model with inverse-square interactions and the twisted $\su(m)$ XXX Heisenberg
model~\cite{ABB88,Ni13,Ga14,NW14}.

We shall study in detail the solvable model with inverse square interactions and ordinary
(minimally polarized) spin flip operators mentioned above, which formally reduces to the HS chain
if the latter operators are replaced by the identity. Taking advantage of the connection between
this chain and its associated spin dynamical model, we shall evaluate the chain's partition
function in closed form. As it turns out, the structure of this partition function depends
critically on the parity of the number $m$ of spin degrees of freedom. In particular, for even $m$
the partition function factorizes as the product of the partition functions of the
(antiferromagnetic) $\su(m/2)$ and $\su(1|1)$-supersymmetric HS chains. This fact is certainly
surprising, since the new model is built from standard (non-supersymmetric) permutation operators,
and deserves further study. On a more concrete level, the formula for the partition function in
the even $m$ case straightforwardly yields a complete description of the spectrum, including the
precise degeneracy of each level, in terms of pairs of $\su(1|1)$ and $\su(m/2)$ motifs, and
establishes the invariance of the model under the direct sum of the Yangians $Y(\gl(1|1))$ and
$Y(\gl(0|m/2))$.

The paper's organization is as follows. The new class of models is introduced in
section~\ref{sec.model}, where we also discuss their characteristic invariance under twisted
translations. In section~\ref{sec.spindm} we show how these models arise from a suitable spin
dynamical model of Calogero--Sutherland type through Polychronakos's freezing trick. Using this
fact, in section~\ref{sec.PF} we derive an exact closed-form expression for the partition function
of the novel spin chain with inverse-square interactions. This expression is used in
section~\ref{sec.motifs} to derive a motif-based description of the spectrum of the latter chain
in the even $m$ case. Section~\ref{sec.specdeg} contains a brief discussion of some statistical
properties of the spectrum of this model, in comparison with related models like the $A_{N-1}$ and
$D_{N}$ HS chains. In section~\ref{sec.concout} we present our conclusions, and briefly discuss
several avenues for further research suggested by our results. The paper ends with four technical
appendices in which we describe in detail the derivation of the spectrum of the spin dynamical
model associated with the new solvable chain, evaluate a sum used in section~\ref{sec.PF} to
determine its partition function, and compute the standard deviation of its spectrum in closed
form.

\section{The models}\label{sec.model}

In this paper we introduce a remarkable new class of $\su(m)$ spin chains with long-range
interactions. We shall start by discussing the simplest representative of the latter class, with
Hamiltonian
\begin{equation}
  \cH = \frac14\sum_{1\le i<j\le N}\bigg[\frac{1+S_{ij}}{\sin^2(\th_i-\th_j)}
  +\frac{1+\tS_{ij}}{\cos^2(\th_i-\th_j)}\bigg]\,,\qquad \th_k:=\frac{k\pi}{2N}\,,
  \label{Hchain}
\end{equation}
where $S_{ij}$ is the operator permuting the $i$-th and $j$-th spins, $\tS_{ij}=S_{ij}S_iS_j$, and
$S_k$ is a local operator at site $k$ satisfying $S_k^2=1$ for all $k$. More precisely, if we
label the $\su(m)$ internal degrees of freedom by the half-integers~$-M,-M+1,\dots,M:=(m-1)/2$,
the action of the operator~$S_{ij}$ on the canonical basis states
\[
  \ket{s_1,\dots,s_N}=:\ket\bbs\,,\qquad s_k\in\{-M,-M+1,\dots,M\}\,,
\]
of the internal (spin) Hilbert space $\Si:=\bigotimes\limits_{i=1}^N\CC^m$ is given by
\[
  S_{ij}\ket{s_1,\dots,s_i,\dots,s_j,\dots,s_N}=\ket{s_1,\dots,s_j,\dots,s_i,\dots,s_N}\,.
\]
The operators $S_{ij}$ and $S_iS_j$ (with $1\le i<j\le N$) generate the Weyl group of the $D_N$
classical root system, algebraically determined by the non-trivial relations
\[
  S_{ij}^2=(S_iS_j)^2=1,\qquad S_{ij}S_{jk}=S_{ik}S_{ij}=S_{jk}S_{ik}, \qquad S_{ij}S_iS_k=S_jS_k
  S_{ij}\,,
\]
where the indices~$i,j,k$ are all distinct and it is understood that $S_{ij}=S_{ji}$, for $i>j$.
The model~\eqref{Hchain}, however, does \emph{not} have the standard form\footnote{From now on,
  unless otherwise stated all summations and products will range over the set~$\{1,\dots,N\}$.}
\begin{equation}\label{cHDN}
  \cH_D=\sum_{i<j}\Big[f(\xi_i-\xi_j)(1+S_{ij})+f(\xi_i+\xi_j)(1+\tS_{ij})\Big]
\end{equation}
of a spin chain with sites~$\xi_k$ associated with the $D_N$ root system~\cite{BFG09,BFG11}. Nor
is it clearly associated with the~$A_N$ root system like, e.g., the original HS chain and its
rational~\cite{Fr93,Po93}, hyperbolic~\cite{FI94} and elliptic~\cite{In90} counterparts, due to
the presence of the operators~$S_i$ in the Hamiltonian. Thus the chain Hamiltonian~\eqref{Hchain}
is not directly associated with an extended root system, unlike most chains of HS type considered
so far in the literature~(see, e.g., \cite{Ya95,BPS95,BFG11,BFG13} for other instances of HS-like
chains of $BC_N$, $B_N$ and $D_N$ types).

For each $k=1,\dots,N$ the operator $S_k$ generates the cyclic group $Z_2$ of order $2$, which has
two inequivalent irreducible representations (necessarily one-dimensional, as $Z_2$ is abelian)
$R_\pm$ for which $S_k=\pm1$. If the internal space $\CC^m $ decomposes under the action of each
operator $S_k$ as $\CC^m=R_+^l\oplus R_-^{m-l}$, we shall say that the degree of polarization of
the corresponding $m$-dimensional representation of $Z_2$ is $d=|2l-m|\in\{\vep,\vep+2,\dots,m\}$
(where $\vep\in\{0,1\}$ is the parity of $m$). When $d=m$ is maximal, i.e., when $S_k$ is either
$1$ or $-1$ for all $k$, $\tS_{ij}=S_{ij}$ and consequently the Hamiltonian~\eqref{Hchain} reduces
to the Hamiltonian~$\cH_{\mathrm{HS}}$ of the HS spin chain
\begin{equation}\label{HSchainA}
  \cH_{\mathrm{HS}}=\frac J2\sum_{i<j}\frac{1+S_{ij}}{\sin^2\Bigl(\tfrac{\pi(i-j)}N\Bigr)}
\end{equation}
with $J=2$. Thus the model~\eqref{Hchain} can be formally regarded as an extension of
the~Haldane--Shastry spin chain. On the other hand, in the remaining cases where the polarization
is not maximal (i.e., for $d <m $), the spin-spin interactions associated with the operators
$S_{ij}$ and $\tilde{S}_{ij}$ do not coincide with each other. As a result, the
Hamiltonian~\eqref{Hchain} describes a novel class of spin chains with uniformly spaced sites (in
fact, we shall see that the sites are arranged on the upper half of a circle). In this paper we
shall focus on the simplest case, with minimal polarization $d=\vep$. It is then convenient to
decompose $\CC^m$ into $(m-\vep)/2$ copies of the two-dimensional (reducible) representation
$R_+\oplus R_-$ plus $\vep$ copies of $R_\pm$. We shall choose the canonical spin basis so that
each space $R_+\oplus R_-$ is spanned by the basis vectors $\{\ket{-i},\ket{i}\}$ with
$i=1,\dots, M$, so that for $\vep=1$ (i.e., if $m$ is odd) the remaining one-dimensional space
$R_\pm$ is spanned by $\ket0$. A moment's thought reveals that we can also choose\footnote{If
  $\vep=1$ and the remaining one-dimensional representation is $R_-$ it suffices to change the
  sign of each $S_k$, which does not change the products $S_iS_j$ appearing in the
  Hamiltonian~\eqref{Hchain}.} the basis so that $S_k\ket i=\ket{-i}$ for $i=1,\dots,M$ and (if
$\vep=1$) $S_k\ket0=\ket0$. Thus in this case $S_k$ acts on the internal space as a spin flip
operator:
\[
  S_k\ket\bbs=\ket{s_1,\dots,-s_k,\dots,s_N}\,.
\]

Except in the trivial case with maximum polarization $d=m$, the two different spin-spin
interactions in the Hamiltonian~\eqref{Hchain} can be regarded as inverse square interactions
between two spins or one spin and its image, as we shall now explain. Indeed, the first term in
eq.~\eqref{Hchain} can be interpreted as a two-body interaction with strength inversely
proportional to the square of the chord distance between the $i$-th and~$j$-th spins, provided
that the chain sites $\xi_k$ are located at the points $\xi_k=\e^{2\iu\th_k}$ in the complex
plane. Since~$\th_k=(k\pi)/(2N)$ with $k=1,\dots,N$, the chain sites lie on uniformly spaced
points in the unit upper \emph{half}-circle~$|z|=1$, $\Im z\ge0$, with angular
coordinates~$\arg \xi_k=2\th_k$. Thus it is natural to regard the model~\eqref{Hchain} as an
\emph{open} chain, in spite of its translationally invariant interactions. If we now define the
\emph{image}~$\xi_k^*$ of the chain site $\xi_k$ as its reflection with respect to the origin,
i.e., $\xi_k^*=-\xi_k$, we have
\[
  \big|\xi_k-\xi_j^*\big|=\big|\e^{2\iu\th_k}+\e^{2\iu\th_j}\big|=2\cos(\th_j-\th_k).
\]
Thus the second term in the Hamiltonian~\eqref{Hchain} can be interpreted as a two-body
interaction of each spin in the chain with the image of any other spin, whose strength is again
inversely proportional to the square of their distance.

We shall next discuss the remarkable behavior of the Hamiltonian~\eqref{Hchain} under translations
along the chain sites, which, as mentioned in the Introduction, is one of the model's distinctive
features. To this end, let $T$ denote the left translation operator defined by
\[
  T\ket{s_1,\dots,s_N}:=\ket{s_2,\dots,s_N,s_1}\,,
\]
so that $T^\dagger=T^{-1}$ and
\[
  T^\dagger S_{ij}T=S_{i+1,j+1},\qquad T^\dagger \tS_{ij}T=\tS_{i+1,j+1}
\]
with
\begin{equation}\label{percond}
  S_{k,N+1}\equiv S_{1k}\,,\qquad\tS_{k,N+1}\equiv\tS_{1k}\,.
\end{equation}
Since the interaction strength between sites $i$ and $j$ in the Hamiltonian~\eqref{Hchain} depends
only on $i-j$, it may naively seem that $\cH$ commutes with the translations generator~$T$ when we
use eqs.~\eqref{percond} to deal with boundary terms. This is actually not the case due to the
open nature of the model (except, of course, in the trivial case $d=m$), which is also apparent
from the fact that the interaction strengths of site $1$ with sites $2$ and $N$ differ. More
precisely, taking eq.~\eqref{percond} into account we obtain
\begin{align*}
  T^\dagger\cH T&=\frac14\sum_{1\le i<j\le N}\bigg[\frac{1+S_{i+1,j+1}}{\sin^2(\th_i-\th_j)}
                  +\frac{1+\tS_{i+1,j+1}}{\cos^2(\th_i-\th_j)}\bigg]
                  =\frac14\sum_{2\le i<j\le N}\bigg[\frac{1+S_{i,j}}{\sin^2(\th_i-\th_j)}
                  +\frac{1+\tS_{i,j}}{\cos^2(\th_i-\th_j)}\bigg]\\
                &+
                  \frac14\sum_{j=2}^{N}\bigg[\frac{1}{\sin^2(\th_j-\th_1)}+\frac{1}{\cos^2(\th_j-\th_1)}+
                  \frac{S_{1j}}{\sin^2(\th_{j-1}-\th_N)}
                  +\frac{\tS_{1j}}{\cos^2(\th_{j-1}-\th_N)}\bigg]\,.
\end{align*}
From the identities
\[
  \sin^2(\th_{j-1}-\th_N)=\sin^2\bigl(\tfrac{\pi(j-1-N)}{2N}\bigr)
  =\cos^2\bigl(\tfrac{\pi(j-1)}{2N}\bigr)\equiv\cos^2(\th_j-\th_1),
\]
and similarly~$\cos^2(\th_{j-1}-\th_N)=\sin^2(\th_j-\th_1)$, it then follows that
\begin{align*}
  T^\dagger\cH T&=\frac14\sum_{2\le i<j\le N}\bigg[\frac{1+S_{ij}}{\sin^2(\th_i-\th_j)}
                  +\frac{1+\tS_{ij}}{\cos^2(\th_i-\th_j)}\bigg]
                  +\frac14\sum_{j=2}^N\bigg[\frac{1+S_{1j}}{\cos^2(\th_{j}-\th_1)}
                  +\frac{1+\tS_{1j}}{\sin^2(\th_{j}-\th_1)}\bigg]\\
                &=\cH-\sum_{j=2}^N\frac{\cos\bigl(2(\th_j-\th_1)\bigr)}{\sin^2\bigl(2(\th_j-\th_1)\bigr)}\,S_{1j}(1-S_1S_j)\,.               
\end{align*}
Thus $\cH$ does not commute with the left translation~$T$ unless $S_j=\pm1$ for all $j$, i.e., in
the trivial case of maximal polarization when the model~\eqref{Hchain} reduces to the HS spin
chain. However, this problem is easily cured introducing the ``twisted'' (left) translation
operator
\[
  \cT:=TS_1\,.
\]
Indeed, taking into account that
\[
  \cT^\dagger S_{ij}\cT=S_1T^\dagger S_{ij}TS_1=S_1S_{i+1,j+1}S_1=
  \begin{cases}
    S_{i+1,j+1}\,,& j<N\\
    S_1S_{1,i+1}S_1=S_1S_{i+1}S_{1,i+1}=\tS_{1,i+1}\,,& j=N\,,
  \end{cases}
\]
and similarly for~$\cT^\dagger \tS_{ij}\cT$ (since $S_{i+1,j+1}$ or $\tS_{i+1,j+1}$ commutes with
$S_1$ unless $j=N$) and proceeding as above we obtain
\begin{align*}
  \cT^\dagger\cH \cT=&\frac14\sum_{2\le i<j\le N}\bigg[\frac{1+S_{ij}}{\sin^2(\th_i-\th_j)}
                       +\frac{1+\tS_{ij}}{\cos^2(\th_i-\th_j)}\bigg]\\
                     &+\frac14\sum_{j=2}^{N}\bigg[\frac{1}{\sin^2(\th_j-\th_1)}+\frac{1}{\cos^2(\th_j-\th_1)}+
                       \frac{\tS_{1j}}{\sin^2(\th_{j-1}-\th_N)}
                       +\frac{S_{1j}}{\cos^2(\th_{j-1}-\th_N)}\bigg]=\cH\,.
\end{align*}
We thus see that the chain Hamiltonian~\eqref{Hchain} commutes with the elements of the group of
twisted translations generated by $\cT$. Since~$S_iT=TS_{i+1}$ we have $\cT^k=T^kS_k\cdots S_1,$
and hence
\[
  \cT^N=S_N\cdots S_1\,.
\]
It follows that $\cT^{2N}=1$, so that the latter group is a cyclic group of order $2N$ (i.e.,
twice that of the group of standard translations). Of course, for $d=m$ (i.e, when the
model~\eqref{Hchain} reduces to the HS chain), and only in this case, the twisted translation
operator reduces to the standard one $T$. This again underscores the fundamentally new character
of the model~\eqref{Hchain} in the case of non-maximal polarization $d<m$. In particular, in the
case of minimal polarization $d=\vep$ the operator~$\cT$ acts on the canonical spin basis as
\[
  \cT\ket{s_1,\dots,s_N}=\ket{s_2,\dots,s_N,-s_1}\,.
\]
Thus in this case, which shall be extensively dealt with in the remaining sections, the twisted
translation consists of a translation followed by a flip of the last site's spin.
\begin{rem}
  The twisted translation operator~$\cT$ satisfies the identities
  \[
    \cT^\dagger S_{ij}\cT=S_{i+1,j+1},\qquad \cT^\dagger \tS_{ij}\cT=\tS_{i+1,j+1}
  \]
  provided that we identify $S_{k,N+1}$ with $\tS_{1k}$ and $\tS_{k,N+1}$ with $S_{1k}$, which is
  akin to using twisted boundary conditions. Note, however, that we have not used twisted boundary
  conditions (or, indeed, boundary conditions of any kind) in the definition of the
  model~\eqref{Hchain}, since its Hamiltonian contains only operators with indices in the range
  $1,\dots,N$. In particular, the new model is not a variant of the twisted HS chain introduced by
  Fukui and Kawakami~\cite{FK96}. Indeed, after a gauge transformation of the original spin
  operators the Hamiltonian of the latter model describes a translationally invariant chain with
  sites uniformly arranged on a circle and periodic boundary conditions, differing from the
  original HS chain only in the strength of the spin-spin interactions.
\end{rem}

In fact, there is a much wider class of models sharing with the chain~\eqref{Hchain} its symmetry
under twisted translations. To see this, consider the Hamiltonian
\[
  \cH = \sum_{1\le i<j\le N}\big[f(i-j)(1+S_{ij})+\tf(i-j)(1+\tS_{ij})\big]\,,
\]
where $f$ is an even function of one variable. A straightforward computation along the previous
lines leads to the identity
\begin{align*}
  \cT^\dagger\cH\cT=&\sum_{2\le i<j\le N}\big[f(i-j)(1+S_{ij})+\tf(i-j)(1+\tS_{ij})\big]\\
                    &+\sum_{j=2}^N\big[f(j-1-N)(1+\tS_{1,j})+\tf(j-1-N)(1+S_{1j})\big]\,.
\end{align*}
We thus see that $\cH$ will commute with $\cT$ provided that the following two identities between
the unknown functions $f$ and $\tf$ are satisfied:
\[
  \tf(x)=f(x-N)\,,\qquad f(x)=\tf(x-N)\,.
\]
The first identity relates $\tf$ to $f$, while the second one is then equivalent to the condition
that $f$ be $2N$-periodic. These considerations lead to the Hamiltonian
\begin{equation}
  \label{cHgen}
  \cH = \sum_{1\le i<j\le N}\big[f(i-j)(1+S_{ij})+f(i-j-N)(1+\tS_{ij})\big]\,,
\end{equation}
with $f$ an even $2N$-periodic function. This periodicity condition makes it natural to regard the
latter model as a uniformly spaced semi-circular (open) chain, with sites
$\xi_k=\e^{\iu k\pi/N}\equiv\e^{2\iu\th_k}$. The model~\eqref{Hchain} is then obtained requiring
that $f(i-j)$ be the square of the inverse distance between sites $i$ and $j$.

\begin{rem}
  The new chain~\eqref{Hchain} is also related to the (non-translationally invariant) HS chain of
  $D_N$ type. To see this, note first of all that the Hamiltonian~\eqref{Hchain} can be written as
  \begin{equation}\label{Hgen}
    \cH=\sum_{1\le i<j\le N}\bigg(\frac{1+S_{ij}}{|\xi_i-\xi_j|^2}
    +\frac{1+\tS_{ij}}{|\xi_i-\xi_j^*|^2}\bigg)\,,
  \end{equation}
  where the chain sites~$\xi_k=\e^{2\iu\th_k}=-\xi_k^*$ are the coordinates of the unique (up to
  an overall translation) equilibrium of the potential
  \begin{equation}\label{Ugen}
    U(\bx)=\sum_{1\le i<j\le N}\bigg(\frac1{|z_i-z_j|^2} +\frac1{|z_i-z_j^*|^2}\bigg),
  \end{equation}
  with~$z_k=-z_k^*=\e^{2\iu x_k}$ in the set $x_1<\cdots<x_N<x_1+\pi/2$. If, on the other hand, we
  define the image~$\xi_k^*$ by~$\xi_k^*=\overline{\xi_k}$ (where the bar denotes complex
  conjugation), then
  \[
    \big|\xi_k-\xi_j^*\big|=\big|\e^{2\iu\th_k}-\e^{-2\iu\th_j}\big|=2\big|\sin(\th_j+\th_k)\big|\,.
  \]
  Taking $\th_k$ as the coordinates of the unique equilibrium (in the
  set~$|x_1|<x_2<\cdots<x_N<\pi-x_{N-1}$) of the potential~\eqref{Ugen} with
  $z_j^*=\overline{z_j\vrule width0pt height 6pt}$, the resulting Hamiltonian~\eqref{Hgen} with
  $\xi_k^*=\overline{\xi_k}$ becomes that of the HS chain of~$D_N$ type\footnote{The sites $\th_i$
    of the HS chain of $D_N$ type are explicitly given by $\th_i=(\arccos x_i)/2$, where
    $x_1,\dots,x_N$ are the zeros of the Jacobi polynomial $P_N^{-1,-1}$~\cite{BFG11}.}
  \begin{equation}\label{HSchainD}
    \cH_{\text{HS,D}}=\frac J2\sum_{i<j}\left[\frac{1+S_{ij}}{\sin^2(\th_i-\th_j)}
      +\frac{1+\tS_{ij}}{\sin^2(\th_i+\th_j)}\right]
  \end{equation}
  with $J=1/2$.
\end{rem}

\section{Spin dynamical model }\label{sec.spindm}

As mentioned above, from now on we shall deal exclusively with the case of minimal polarization
$d=\vep$, when the operators $S_i$ act simply as ordinary spin flip operators. One of the main
aims of this paper is to derive a closed-form expression for the partition function of the
Hamiltonian~\eqref{Hchain} by exploiting its connection with a non-standard spin dynamical model
introduced in ref.~\cite{FGGRZ01b}.
The existence of such an expression shows that the model~\eqref{Hchain} is a simple but
non-trivial example of an exactly solvable spin chain with long-range two-body interactions and
twisted translation invariance. Note in this respect that the rational and hyperbolic chains of HS
type mentioned above are solvable but not translationally invariant, while the Inozemtsev
chain~\cite{In90} (an HS-like spin chain with elliptic two-body interactions) is translationally
invariant but not solvable.

More precisely, the solvability of the spin chain~\eqref{Hchain} relies on its connection with the
solvable spin dynamical model with Hamiltonian
\begin{align}
  H&=-\De+8a\sum_{1\le i<j\le N}\bigg(\frac{a+S_{ij}}{|z_i-z_j|^2}
     +\frac{a+\tS_{ij}}{|z_i-z_j^*|^2}\bigg)\notag\\
   &=-\De+2a\sum_{i<j}\bigg[\frac{a+S_{ij}}{\sin^2(x_i-x_j)}
     +\frac{a+\tS_{ij}}{\cos^2(x_i-x_j)}\bigg]\,,
     \qquad z_k=-z_k^*:=\e^{2\iu x_k}\,,
     \label{Hspin}
\end{align}
where $\De:=\sum_i\frac{\pd^2}{\pd x_i^2}$ and $a>1/2$. The key idea, which is originally due to
Polychronakos~\cite{Po94}, is to take the large coupling constant limit~$a\to\infty$ in
eq.~\eqref{Hspin}. In this limit the eigenfunctions of $H$ become sharply peaked around the
coordinates of the equilibrium position of the scalar potential~\eqref{Ugen},
explicitly given by
\[
  U(\bx)=\sum_{1\le i<j\le N}\bigg(\frac1{|z_i-z_j|^2} +\frac1{|z_i-z_j^*|^2}\bigg)=
  \sum_{i<j}\sin^{-2}\bigl(2(x_i-x_j)\bigr)\,,
\]
in the configuration space~$A$ of the model~\eqref{Hspin}. Note that this configuration space can
be taken as
\begin{equation}\label{Adef}
  A=\Big\{\bx:=(x_1,\dots,x_N)\in\RR^N:x_1<x_2\dots< x_N<x_1+\tfrac\pi2\Big\}\,,
\end{equation}
due to the impenetrable nature of the singularities of the scalar potential~\eqref{Ugen} on the
hyperplanes $x_i-x_j=k\pi,(2k+1)\pi/2$ with $k\in\ZZ$. The unique global minimum of $U$ in the
open set $A$ is the point whose coordinates~$x_k=k\pi/(2N)\equiv\th_k$ coincide with the sites of
the chain~\eqref{Hchain}, apart from an irrelevant overall translation~\cite{FG14}. Thus in the
limit~$a\to\infty$ the particles' dynamic and spin degrees of freedom effectively decouple. In
fact, since
\[
  H=\Hsc+8a \Hsp(\bx)
\]
with
\begin{align}
  \Hsc&:=-\De+8a(a-1)\sum_{i<j}\sin^{-2}\bigl(2(x_i-x_j)\bigr),\label{Hsc}\\
  \Hsp(\bx)&:=\frac14\sum_{i<j}\bigg[\frac{1+S_{ij}}{\sin^2(x_i-x_j)}
             +\frac{1+\tS_{ij}}{\cos^2(x_i-x_j)}\bigg]\label{Hspinx}
\end{align}
and $\Hsp(\th_1,\dots,\th_N)=\cH$, for~$a\to\infty$ the energies of the spin dynamical
model~\eqref{Hspin} are approximately given by
\[
  E_{ij}\simeq E_i+8a\cE_j\,,
\]
where~$E_i$ and~$\cE_j$ respectively denote two arbitrary eigenvalues of~$\Hsc$ and~$\cH$. From
the above relation we immediately obtain the following \emph{exact} formula for the partition
function~$\cZ$ of the chain~\eqref{Hchain}:
\begin{equation}
  \label{Zft}
  \cZ(T)=\lim_{a\to\infty}\frac{Z(8aT)}{\Zsc(8aT)}\,,
\end{equation}
where~$Z$ and~$\Zsc$ are respectively the partition functions of the Hamiltonians~$H$ and~$\Hsc$.

In fact, the general spin chain Hamiltonian~\eqref{cHgen} can also be derived from a spin
dynamical model akin to eq.~\eqref{Hspin}. Indeed, let
\begin{equation}\label{HgenF}
  H=-\De+2a\sum_{i<j}\big[F(x_i-x_j)(a+S_{ij})+F(x_i-x_j-\tau)(a+\tS_{ij})]\,,
\end{equation}
where $F(x)$ is an even $2\tau$-periodic function\footnote{To obtain a purely discrete spectrum,
  we also need that $F(x)>0$ for $0<x<2\tau$ and $F(x)\to\infty$ as $x\to0+$.}. The spin chain
Hamiltonian constructed from $H$ applying the freezing trick discussed above is given (up to a
proportionality factor which can be chosen at will) by
\begin{equation}\label{HchainF}
  \cH=\sum_{i<j}\big[F(\xi_j-\xi_i)(1+S_{ij})+F(\xi_j-\xi_i-\tau)(1+\tS_{ij})]\,,
\end{equation}
where $(\xi_1,\dots,\xi_N)=:\bxi$ are the coordinates of the equilibrium of the scalar potential
\[
  U(\bx)=\sum_{i<j}u(x_i-x_j)\,,\qquad u(x):=F(x)+F(x-\tau)\,,
\]
in the region $\{\bx\in\RR^N:x_1<\cdots <x_N<x_1+\tau\}$. To see that $\bxi$ is uniquely defined
(up to a rigid translation), note first of all that $u$ is even and $\tau$-periodic. Indeed,
\[
  \left\{
    \begin{aligned}
      u(-x)&=F(-x)+F(-x-\tau)=F(x)+F(x+\tau)=F(x)+F(x-\tau)\equiv u(x)\,,\\
      u(x+\tau)&=F(x+\tau)+F(x)=F(x-\tau)+F(x)\equiv u(x)\,.
    \end{aligned}
  \right.
\]
It then follows that $u'$ is odd and $\tau$-periodic. Furthermore, $u'(\tau/2)=0$, since
\[
  u'(x+\tau)=u'(x)\implies u'(\tau/2)=u'(-\tau/2)=-u'(\tau/2)\,.
\]
Reasoning as in Theorems 1-2 of ref.~\cite{FG14} we conclude that
\[
  \xi_k=\frac{k\tau}N\,,\qquad k=1,\dots,N\,,
\]
provided that $u''(x)=F''(x)+F''(x-\tau)>0$ for $0<x<\tau$. Thus the chain~\eqref{HchainF}
associated with the Hamiltonian~\eqref{HgenF} coincides with~\eqref{cHgen} (up to a
proportionality factor) if we set
\[
  f(k)=F(k\tau/N)\,.
\]
For instance, taking $F(x)=\sin^{-2}x$ we have $2\tau=\pi$ and $f(k)=\sin^{-2}(k\pi/2N)$, yielding
the model~\eqref{Hchain}. Another interesting possibility is $F(x)=\sn^{-2}(x;k)\equiv\sn^{-2}x$,
so that $\tau=K(k)$ (the complete elliptic integral of the first kind),
\begin{equation}\label{Hell}
  H=-\De+2a\sum_{i<j}\bigg[\frac{a+S_{ij}}{\sn^2(x_i-x_j)}+\frac{\dn^2(x_i-x_j)}{\cn^2(x_i-x_j)}\,
  (a+\tS_{ij})\bigg]\,,
\end{equation}
and
\begin{equation}\label{cHell}
  \cH=\sum_{i<j}\bigg[\frac{1+S_{ij}}{\sn^2(\xi_i-\xi_j)}
  +\frac{\dn^2(\xi_i-\xi_j)}{\cn^2(\xi_i-\xi_j)}\, (1+\tS_{ij})\bigg]\,,\qquad
  \xi_j:=\frac{jK(k)}N\,.
\end{equation}
These elliptic models (or their equivalent counterparts constructed using Weierstrass elliptic
functions) appear to be new and deserve further study in their own right. In fact, it should be
noted that the Hamiltonian~\eqref{Hell} does not reduce in the maximally polarized case $S_i=1$ to
the standard elliptic model $-\De+8a\sum_{i<j}\sn^{-2}\bigl(2(x_i-x_j)\bigr)\,(a+S_{ij})$, whose
associated spin chain is the Inozemtsev chain.

\begin{rem}
  The dynamical model~\eqref{Hell} and the spin chain~\eqref{cHell} clearly reduce to the
  model~\eqref{Hspin} and the chain~\eqref{Hchain} for $k=0$. Since the Inozemtsev elliptic chain
  yields the Heisenberg XXX (nearest neighbors) model in the limit when the imaginary period of
  the Weierstrass function tends to zero, it is of interest to investigate the analogous limit of
  the spin chain~\eqref{cHell}. In fact, in order to make this limit well defined we shall
  consider instead the variant of the elliptic chain~\eqref{cHell} obtained by taking
  \begin{equation}\label{fsn}
    f(x)=\frac{K'^2}{\pi^2}\,\e^{\frac{\pi
        K}{NK'}}\bigg\{\sn^{-2}\bigl(\tfrac{Kx}N\bigr)-\frac13\,(1+k^2)
    -\frac{\pi^2}{12K'^2}\bigg\}
  \end{equation}
  instead of $\sn^{-2}(Kx/N)$ in eq.~\eqref{cHgen}, where $K'\equiv K'(k):=K(\sqrt{1-k^2}\,)$.
  Taking into account the identity~\cite{La89}
  \[
    \sn^{-2}z=\wp(x;K,\iu K')+\frac13\,(1+k^2)\,,
  \]
  where $\wp(z;\om_1,\om_3)$ denotes the Weierstrass elliptic function with half-periods
  $(\om_1,\om_3)$, we can equivalently write
  \begin{equation}\label{fwp}
    f(x)=\e^{\frac{\pi
        K}{NK'}}\bigg(\frac{K'^2}{\pi^2}\,\wp\bigl(\tfrac{Kx}N;K,\iu K'\bigr)
    -\frac1{12}\bigg)=\e^{\frac{2\pi}\al}\bigg(\frac{\al^2}{4\pi^2}\,\wp_{2N}(x)
    -\frac1{12}\bigg)
  \end{equation}
  where $\al:=\frac{2NK'}{K}$ and $\wp_{2N}(x):=\wp(x;N,\iu\al/2)$ is a Weierstrass function with
  real period $2N$. As $k\to1$, the elliptic integrals $K(k)$ and $K'(k)$ tend respectively to
  $\infty$ and $\pi/2$, so that $\al\to0$. Thus if $1\le x\le 2N-1$ we have
  \[
    \lim_{k\to1}f(x)=\lim_{\al\to0}\e^{\frac{2\pi}\al}\bigg(\frac{\al^2}{4\pi^2}\,\wp_{2N}(x)
    -\frac1{12}\bigg)=\de_{1,x}+\de_{2N-1,x}
  \]
  (see, e.g., ref.~\cite{FG14JSTAT}, for a detailed proof of the latter identity). In particular,
  if $1\le x\le N-1$ then $1\le N-x\le N-1$ as well, and hence
  \[
    \lim_{k\to1}f(x-N)=\lim_{k\to1}f(N-x)=\de_{1,N-x}+\de_{2N-1,N-x}=\de_{N-1,x}\,.
  \]
  Since $\de_{2N-1,x}=0$ for $1\le x\le N-1$, the $k\to1$ limit of the chain~\eqref{cHgen} with
  $f$ given by eq.~\eqref{fsn} or \eqref{fwp} is the homogeneous near-neighbors chain
  \[
    \cH=\sum_{i=1}^{N-1}(1+S_{i,i+1})+1+\tS_{1N}\,,
  \]
  which we can rewrite as
  \begin{equation}\label{Hlim}
    \cH=\sum_{i=1}^{N}(1+S_{i,i+1})
  \end{equation}
  provided that we set
  \begin{equation}\label{twist}
    S_{N,N+1}\equiv\tS_{1N}.
  \end{equation}
  For $m=2$ we have
  \[
    S_{ij}=\frac12\,(1+\bsi_i\cdot\bsi_j)\,,\qquad S_i=\si_i^x\,,
  \]
  where $\si_k^\al$ (with $\al=x,y,z$) is the Pauli matrix $\si^\al$ acting on the $k$-th site and
  $\bsi_k:=(\si_k^x,\si_k^y,\si_k^z)$. Moreover, in this case the twisted boundary
  condition~\eqref{twist} simplifies to
  \begin{multline*}
    \frac12(1+\bsi_{N}\cdot\bsi_{N+1})\equiv \frac12(1+\bsi_{1}\cdot\bsi_{N})\si_1^x\si_N^x
    =\frac12(\si_1^x\si_N^x+\bsi_{1}\si_1^x\cdot\bsi_{N}\si_N^x)\\
    =\frac12\Big[\si_1^x\si_N^x+(1,-\iu\si_1^z,\iu\si_1^y)\cdot(1,-\iu\si_N^z,\iu\si_N^z)]
    =\frac12(1+\si_1^x\si_N^x-\si_1^y\si_N^y-\si_1^z\si_N^z)\,,
  \end{multline*}
  which is equivalent to setting
  \[
    \si_{N+1}^x\equiv\si_1^x\,,\qquad \si_{N+1}^{y,z}\equiv-\si_1^{y,z}\,.
  \]
  Hence for $m=2$ the chain~\eqref{Hlim} is the Heisenberg XXX model with anti-periodic boundary
  conditions~\cite{ABB88,Ni13,Ga14,NW14}. Note, finally, that proceeding along the same lines as
  in ref.~\cite{FG14JSTAT} it is not difficult to construct a spin chain Hamiltonian of the
  form~\eqref{cHgen} depending on a parameter that smoothly interpolates between the trigonometric
  model~\eqref{Hchain} and the twisted $\su(m)$ XXX Heisenberg chain~\eqref{Hlim}. Indeed, it
  suffices to take
  \begin{equation}\label{interp}
    f(x)=\frac{\al^2}{\pi^2}\sinh^2\bigl(\tfrac\pi\al\bigr)\biggl(\wp\bigl(x;N,\tfrac{\iu\al}2\bigr)
    -\frac2{\pi^2}\,\eta_1\bigr(\tfrac12,\tfrac{\iu N}{\al}\bigr)\biggr)\,,
  \end{equation}
  where $\al>0$, $\eta_1(\om_1,\om_3):=\ze(\om_1;\om_1,\om_3)$ and $\zeta(z;\om_1,\om_3)$ denotes
  the Weierstrass zeta function satisfying $\ze'(z;\om_1,\om_3)=\wp(z;\om_1,\om_3)$. It is then
  straightforward to show that the $\al\to\infty$ limit of the chain~\eqref{cHgen}-\eqref{interp}
  is the Hamiltonian~\eqref{Hchain} multiplied by $\pi^2/N^2$, whereas its $\al\to0$ limit is the
  twisted Heisenberg chain~\eqref{Hlim}.
\end{rem}

\section{Partition function}\label{sec.PF}

In this section we apply the freezing trick formula~\eqref{Zft} to evaluate in closed form the
partition function of the spin chain~\eqref{Hchain}. In fact, for the scalar model $\Hsc$ in
eq.~\eqref{Hsc} it suffices to note that the change of variables $y_k=2x_k$ transforms~$\Hsc$ into
$4H_S$, where
\[
  H_S:=-\sum_{i}\frac{\pd^2}{\pd y_i^2}+2a(a-1)\sum_{i<j}\sin^2(y_i-y_j)
\]
is the Hamiltonian of the scalar Sutherland model~\cite{Su71b,Su71}. From the formula in
ref.~\cite{FG05} for the spectrum of the latter model in the center of mass (CM) frame it follows
that the spectrum of $\Hsc$ in the CM frame is given by
\[
 E_{\bn}=4\sum_i\Big[2p_i+a(N-2i+1)\Big]^2\,;\qquad
  |\bn|:=\sum_in_i,\quad p_i:= n_i-\frac{|\bn|}N\,,
\]
with $n_k\in\mathbb Z$ and $\ds n_1\geqslant\dots\geqslant n_{N-1}\geqslant n_N=0$. Expanding in
powers of~$a$ the energies $E_{\bn}$ in the latter equation we obtain
\[
  E_{\bn}=E_0+16a\sum_ip_i(N-2i+1)+O(1),
\]
where
\[
  E_0=4a^2\sum_i(N+1-2i)^2=\frac43\,N(N^2-1)a^2
\]
is the ground-state energy of~$\Hsc$ in the CM frame. We thus have
\begin{equation}\label{Zsc}
  \lim_{a\to\infty}q^{-\frac{E_0}{8a}}\Zsc(8aT)=\sum_{n_1\ge\cdots\ge n_{N-1}\ge0}q^{2\sum_ip_i(N+1-2i)}
  =\prod_{i=1}^{N-1}(1-q^{2i(N-i)})\,,\qquad q:=\e^{-1/(k_BT)},
\end{equation}
where the sum was evaluated in ref.~\cite{FG05}.

We need next to compute the spectrum of the spin Hamiltonian~$H$ in the CM frame. This is easily
achieved by boosting its eigenfunctions in the Hilbert space~$\fH=\La_a\La_0(\fH'\otimes\Si)$,
computed in appendix~\ref{app.specHscH}, to the CM frame, and evaluating their resulting energies.
To this end, it suffices to note that if a state $\Psi\in\fH'\otimes\Sigma$ is a simultaneous
eigenfunction of $H$ and $P$ with respective eigenvalues $E$ and $p$,
then~$\e^{\iu\ms c\ms|\bx|}\Psi$ (where~$|\bx|:=\sum_kx_k$) is an eigenfunction of these operators
with eigenvalues~$E+2cp+Nc^2$ and $p+Nc$. In particular, $\e^{-\frac{\iu p}{N}\ms|\bx|}\Psi$ is an
eigenfunction of $H$ with zero momentum and energy $E-\frac{p^2}N$\,. As we saw in
appendix~\ref{app.specHscH}, the eigenfunctions of $H$ in $\fH$ are of the form
\[
  \Psi^\vep_{\bk,\bbs}=\La_a\La_0^\vep(\psi_\bk(\bx)\ket\bbs)\,,
\]
where $\vep=\pm1$ and the quantum numbers~$\bk\in\ZZ^N$, $\bbs\in{-M,-M+1,\dots,M}$ satisfy
conditions 1--3 in the latter appendix (in particular, $k_1\ge\cdots\ge k_N$). Setting
$l=k_n\in\ZZ$ and $\bn=\bk-l(1,\dots,1)$, and taking into account eq.~\eqref{boost}, we have
\[
  \Psi^\vep_{\bk,\bbs}=\La_a\La_0^\vep(\e^{2\iu l|\bx|}\psi_\bn(\bx)\ket\bbs) =\e^{2\ms\iu\ms
    l\ms|\bx|}\La_a\La_0^{(-1)^l\vep}\big(\psi_{\bn}\ket\bbs\big)
  =\e^{2\ms\iu\ms
    l\ms|\bx|}\Psi^{(-1)^l\vep}_{\bn,\bbs}\,,
\]
with $n_1\ge\cdots\ge n_N=0$. Since the factor $\e^{2\ms\iu\ms l\ms|\bx|}$ is just a momentum
boost, the wave functions $\Psi^\vep_{\bk,\bbs}$ and $\Psi^{(-1)^l\vep}_{\bn,\bbs}$ should be
regarded to represent the same physical state in two different reference frames. Thus we need only
consider eigenfunctions of~$H$ of the form $ \Psi^\pm_{\bn,\bbs}$ with $n_1\ge\cdots\ge n_N=0$,
whose energy and momentum are respectively
\[
  E(\bn)=4\sum_i\Big[n_i+a(N-2i+1)\Big]
\]
and $2|\bn|$ (cf. appendix~\ref{app.specHscH}). As a consequence, the boosted states
\begin{equation}\label{Hpsi}
  \e^{-\frac{2\iu|\bn||\bx|}N}\Psi_{\bn,\bbs}^\pm
  =\e^{-\frac{2\iu|\bn||\bx|}N}\La_a\La_0^{\pm}(\psi_{\bn}\ket\bbs)\,,\qquad n_1\ge\cdots\ge n_N=0\,,
\end{equation}
where the spin quantum numbers~$\bbs$ satisfies conditions 2--3 in the latter appendix, are a
basis of eigenfunctions of~$H$ in the CM frame with corresponding energies
\begin{equation}\label{HspecCM}
  E_{\bn,\bbs}^\pm:= E(\bn)-\frac{4|\bn|^2}{N}=4\sum_i\Big[p_i+a(N-2i+1)\Big]^2\,,\qquad
  p_i=n_i-\frac{|\bn|}N\,.
\end{equation}
\begin{rem}\label{rem.four}
  Remarkably, from the previous equation and condition 3 in appendix~\ref{app.specpiper} it
  follows that in the case of even $m$ the eigenvalues of~$H$ coincide with those
  of~$H_S^{(m/2)}(2a)$, where
  \begin{equation}\label{Ha}
    H_S^{(m/2)}(a)=-\De+2a\sum_{i<j}\frac{a+S_{ij}}{\sin^2(x_i-x_j)}
  \end{equation}
  is the Hamiltonian of the~$\su(m/2)$ spin Sutherland model~\cite{HH92}, but with \emph{twice}
  the degeneracy for each level.
\end{rem}

\medskip
We are now ready to evaluate the partition function of the spin dynamical model~\eqref{Hspin} in
the large $a$ limit. As it turns out, the calculation depends in an essential way on the parity of
the number~$m$ of internal degrees of freedom.

\subsection{Even~$m$}

Let us start by parametrizing the multiindex~$\bn$ in eq.~\eqref{HspecCM} as
\begin{equation}\label{bnnu}
  \bn=\big(\underbrace{\nu_1,\dots,\nu_1}_{\ell_1}\,,\dots,
  \underbrace{\nu_{r-1},\dots,\nu_{r-1}}_{\ell_{r-1}}\,,\underbrace{0,\dots,0}_{\ell_r}\big)\,,
\end{equation}
where $\nu_1>\cdots>\nu_{r-1}>0$ (with~$\nu_k\in\NN$) and $\ell_1+\cdots +\ell_r=N$
(with~$\ell_i>0$ for all $i$). Since in this case the components~$s_i$ of the spin vector~$\bbs$
in eq.~\eqref{HspecCM} cannot take the value zero, condition~3 in appendix~\ref{app.specpiper}
simply states that $s_i$ can only take the $m/2$ positive values $1/2,3/2,\dots, M=(m-1)/2$. By
condition 2, there are $\binom{m/2}{\ell_i}$ possible choices of the components of~$\bbs$
corresponding to the~$\ell_i$ components of the vector~$\bn$ equal to~$\nu_i$. Since the
energies~$E^\vep_{\bn,\bbs}$ do not depend on the quantum numbers~$\vep\in\{\pm1\}$ and~$\bbs$,
for each~$\bn$ they have thus an intrinsic degeneracy given by
\[
  2\prod_{i=1}^r\binom{m/2}{\ell_i}=:2d(\bell)\,.
\]
Expanding~$E_{\bn,\bbs}^\vep$ in powers of~$a$ we obtain
\[
  E_{\bn,\bbs}^\vep=E_0+8a\sum_ip_i(N-2i+1)+O(1)\,,
\]
and therefore
\begin{equation}\label{Z8aTstart}
  \lim_{a\to\infty}q^{-\frac{E_0}{8a}}Z(8aT)=2\sum_{\bell\in\mP_N}d(\bell)\sum_{\nu_1>\cdots>
    \nu_{r-1}>0}q^{\sum_ip_i(N+1-2i)}\,,
\end{equation}
where~$\mP_N$ is the set of compositions (i.e., ordered partitions) of the integer~$N$. The
sum~$\sum_ip_i(N+1-2i)$ in the latter equation can be evaluated as in ref.~\cite{FG05}, with the
result
\begin{equation}\label{sum}
  \sum_ip_i(N+1-2i)=\sum_{i=1}^{r-1}\tnu_i\cE(L_i)\,,\qquad
  L_i:=\sum_{j=1}^i\ell_j\,,
\end{equation}
where~$\tnu_i:=\nu_i-\nu_{i+1}>0$ and the dispersion relation~$\cE$ is defined by
\[
  \cE(i)=i(N-i)
\]
(see Appendix~\ref{app.sum} for the details). Since the positive integers~$\tnu_i=1,2,\dots$ are
unconstrained, from eq.~\eqref{sum} we easily obtain
\[
  \sum_{\nu_1>\cdots> \nu_{r-1}>0}q^{\sum_ip_i(N+1-2i)}=\sum_{\tnu_1,\dots,\tnu_{r-1}=1}^\infty
  \prod_{i=1}^{r-1}q^{\tnu_i\cE(L_i)} =\prod_{i=1}^{r-1}\sum_{\tnu_i=1}^\infty q^{\tnu_i\cE(L_i)}
  =\prod_{i=1}^{r-1}\frac{q^{\cE(L_i)}}{1-q^{\cE(L_i)}}\,,
\]
and therefore
\begin{equation}\label{Z8aT}
  \lim_{a\to\infty}q^{-\frac{E_0}{8a}}Z(8aT)=2\sum_{\bell\in\mP_N}d(\bell)
  \prod_{i=1}^{r-1}\frac{q^{\cE(L_i)}}{1-q^{\cE(L_i)}}\,.
\end{equation}
Alternatively, this formula can be deduced by exploiting the connection between the spectra of $H$
and the Hamiltonian of the spin Sutherland model described in remark~\ref{rem.four}. Indeed, from
this connection it follows that
\[
  q^{-\frac{E_0}{8a}}Z(8aT)=2q^{-\frac{E_{0,S}(2a)}{8a}}Z_S^{(m/2)}(8aT;2a)=:2f(2a),
\]
where $Z_S^{(m/2)}(T;a)$ is the partition function of the Hamiltonian~\eqref{Ha} and $E_{0,S}(a)$
its ground energy. The $a\to\infty$ of the function $f(a)$ was evaluated in ref.~\cite{FG05},
obtaining as a result the sum in eq.~\eqref{Z8aT}.

Using~eqs.~\eqref{Zsc} and~\eqref{Z8aT} in the freezing trick formula~\eqref{Zft} we finally
obtain the following explicit formula for the partition function of the chain~\eqref{Hchain} in
the case of even~$m$:
\begin{equation}\label{Zeven}
  Z=\prod_{i=1}^N\Big(1+q^{\cE(i)}\Big)
  \cdot\sum_{\bell\in\mP_N} \prod_{i=1}^r\binom{m/2}{\ell_i}\cdot
  q^{\sum\limits_{i=1}^{r-1}\cE(L_i)}
  \cdot\prod_{i=1}^{N-r}\Big(1-q^{\cE(L_i')}\Big)\qquad(\text{even } m)\,,
\end{equation}
with
\[
  \{L_1',\dots,L_{N-r}'\}=\{1,\dots,N-1\}\setminus\{L_1,\dots,L_{r-1}\}.
\]
Remarkably, the latter formula can be written as
\begin{equation}\label{fact}
  Z=Z_{\mathrm{HS}}^{(1|1)}Z_{\mathrm{HS}}^{(m/2)},
\end{equation}
where $Z_{\mathrm{HS}}^{(1|1)}$ and $Z_{\mathrm{HS}}^{(m/2)}$ respectively denote the partition
functions of the antiferromagnetic\footnote{In fact, the spectra of the ferromagnetic and
  antiferromagnetic $\su(1|1)$ HS chains are identical.} $\su(1|1)$ and~$\su(m/2)$ HS spin
chains~\cite{FG05,BB06}. In other words, the model~\eqref{Hchain} with an even number~$m$ of
internal degrees of freedom is isomorphic to the sum of an $\su(1|1)$ and an
antiferromagnetic~$\su(m/2)$ HS chain Hamiltonian acting on the tensor product of their respective
Hilbert spaces. As we shall see in the next section this property, which is far from obvious from
the definition~\eqref{Hchain} of the Hamiltonian, makes it possible to describe the model's
spectrum in terms of Haldane motifs and their corresponding Young tableaux. In particular, in the
simplest $m=2$ case the sum in eq.~\eqref{Zeven} contains only the term with~$\bell=(1,\dots,1)$,
so that
\[
  Z=Z_{\mathrm{HS}}^{(1|1)}q^{\sum\limits_{i=1}^{N-1}\cE(i)}=q^{\frac
    N6\,(N^2-1)}Z_{\mathrm{HS}}^{(1|1)} \qquad(m=2)\,.
\]
Thus the spin~$1/2$ chain~\eqref{Hchain} is equivalent to the~$\su(1|1)$ supersymmetric HS chain,
up to the overall energy shift~$N(N^2-1)/6$. This is again remarkable, since in principle there is
no connection between the spin flip operators~$S_i$ appearing in the Hamiltonian~\eqref{Hchain}
and the~$\su(1|1)$ supersymmetric permutation operator~$S_{ij}^{(1|1)}$.

\subsection{Odd $m$}

According to condition~3 in appendix~\ref{app.specpiper}, when~$m$ is odd the components of the
spin vector~$s_i$ can take the value~$0$ only if the corresponding component~$n_i$ of the
multiindex~$\bn$ has parity~$(1-\vep)/2$. Hence in this case $s_i$ can take~$(m+(-1)^{n_i}\vep)/2$
values instead of~$m/2$, and the intrinsic degeneracy of each energy~\eqref{HspecCM} is thus given
by
\begin{equation*}
  d(\bell,\bnu)
  :=\prod_{i=1}^r\binom{\frac12(m-(-1)^{\nu_i})}{\ell_i}+
  \prod_{i=1}^r\binom{\frac12(m+(-1)^{\nu_i})}{\ell_i}% \\
\end{equation*}
As a consequence, eq.~\eqref{Z8aTstart} now becomes
\begin{equation}\label{Zoddfirst}
  \lim_{a\to\infty}q^{-\frac{E_0}{8a}}Z(8aT)=\sum_{\bell\in\mP_N}\sum_{\nu_1>\cdots>
    \nu_{r-1}>0}d(\bell,\bnu)\,q^{\sum_ip_i(N+1-2i)}\,.
\end{equation}
Introducing again the unconstrained variables~$\tnu_i=\nu_i-\nu_{i+1}\in\NN$ (with
$i=1,\dots,r-1$), in terms of which $\nu_i=\sum_{j=i}^{r-1}\tnu_j$, and using eq.~\eqref{sum} we
obtain
\[
  \lim_{a\to\infty}q^{-\frac{E_0}{8a}}Z(8aT)=\sum_{\bell\in\mP_N}\sum_{\tnu_1,\dots,\tnu_{r-1}=1}^\infty
  d(\bell,\bnu)q^{\sum_{i=1}^{r-1}\tnu_i\cE(L_i)}\,.
\]
The above expression can be further simplified by setting
\[
  \tnu_i=2\tn_i-\de_i\,,\qquad 1\le i\le r-1
\]
where~$\de_i\in\{0,1\}$ is the parity of~$\tnu_i$ and $\tn_i\in\ZZ$ with~$\tn_i\ge1$. Since
$\nu_i=2\sum_{j=i}^{r-1}\tn_j-\sum_{j=i}^{r-1}\de_j$ for $i=1,\dots,r-1$ we have
\begin{multline}\label{degodd}
  d(\bell,\bnu)=\binom{\frac12(m-1)}{\ell_r}\prod_{i=1}^{r-1}\binom{\frac12(m-(-1)^{\De_i})}{\ell_i}\\+
  \binom{\frac12(m+1)}{\ell_r}\prod_{i=1}^{r-1}\binom{\frac12(m+(-1)^{\De_i})}{\ell_i}=:
  d_{\bde}(\bell)\,,\qquad \De_i:=\sum_{j=i}^{r-1}\de_j\,,
\end{multline}
and hence eq.~\eqref{Zoddfirst} is equivalent to
\begin{align*}
  \lim_{a\to\infty}q^{-\frac{E_0}{8a}}Z(8aT)
  &=\sum_{\bell\in\mP_N}\sum_{\de_1,\dots,\de_{r-1}=0}^1
    d_{\bde}(\bell)\sum_{\tn_1,\dots,\tn_{r-1}=1}^\infty
    \prod_{i=1}^{r-1}q^{(2\tn_i-\de_i)\cE(L_i)}\\
  &=\sum_{\bell\in\mP_N}\sum_{\de_1,\dots,\de_{r-1}=0}^1 d_{\bde}(\bell)\prod_{i=1}^{r-1}
    \sum_{\tn_i=1}^\infty q^{(2\tn_i-\de_i)\cE(L_i)}\\
  &=\sum_{\bell\in\mP_N}\sum_{\de_1,\dots,\de_{r-1}=0}^1 d_{\bde}(\bell)\prod_{i=1}^{r-1}
    \frac{q^{(2-\de_i)\cE(L_i)}}{1-q^{2\cE(L_i)}}\,.
\end{align*}
Using eq.~\eqref{Zsc} and the freezing trick relation~\eqref{Zft} we finally obtain the following
explicit formula for the partition function of the chain~\eqref{Hchain} in the odd $m$ case:
\begin{equation}
  \label{Zodd}
  Z=\sum_{\bell\in\mP_N}\prod_{i=1}^{N-r}\big(1-q^{2\cE(L_i')}\big)
  \sum_{\de_1,\dots,\de_{r-1}=0}^1 d_{\bde}(\bell)\,q^{\sum_{i=1}^{r-1}(2-\de_i)\cE(L_i)}
  \qquad(\text{odd }m),
\end{equation}
with~$d_{\bde}(\bell)$ defined by eq.~\eqref{degodd}.

\section{Motifs}\label{sec.motifs}

\subsection{Review of $\su(k|n)$ motifs}
For even $m$, the connection between the partition functions of the $\su(m)$ chain~\eqref{Hchain}
with the antiferromagnetic $\su(1|1)$ and $\su(m/2)$ HS chains embodied in eq.~\eqref{fact} makes
it possible to express the spectrum of the former model in terms of suitable Haldane motifs, as we
shall next explain in detail. To this end, we shall first recall a few essential facts concerning
the description of the spectrum of the $\su(k|n)$ HS chain using supersymmetric motifs.

To begin with, we define the Hamiltonian of the $\su(k|n)$ HS chain
as
\begin{equation}\label{suknHS}
  \cH^{(k|n)}_{\mathrm{HS}}=\frac12\,\sum_{i<j}
  \frac{1-S^{(k|n)}_{ij}}{\sin^2\Bigl(\tfrac{\pi(i-j)}N\Bigr)}\,,
\end{equation}
where $S^{(k|n)}_{ij}$ is the standard $\su(k|n)$ supersymmetric spin permutation operator. More
precisely, the canonical basis of the Hilbert space of $\cH^{(k|n)}$ consists of the $(k+n)^N$
states
\[
  \ket{s_1,\dots,s_N}:=\ket{s_1}\otimes\cdots\otimes\ket{s_N}\,,
\]
where now $s_i\in\{1,\dots,k+n\}$ and we regard as bosonic (resp.~fermionic) the first $k$
(respectively last $n$) internal degrees of freedom. The operator $S_{ij}^{(k|n)}$ is then
defined\footnote{This definition of the supersymmetric permutation operators is equivalent to that
  of Haldane~\cite{Ha93}; see, e.g., refs.~\cite{Ba96,Ba99}.} by
\[
  S_{ij}^{(k|n)}\ket{s_1,\dots,s_i,\dots,s_j,\dots,s_N}
  =(-1)^{n_{ij}(\bbs)}\ket{s_1,\dots,s_j,\dots,s_i,\dots,s_N}\,,
\]
where $n_{ij}(\bbs)$ is $0$ (resp.~$1$) if $s_i$ and $s_j$ are both bosonic (resp.~fermionic), and
is otherwise equal to the number of fermionic spin variables $s_k$ in $\ket{s_1,\dots,s_N}$ with
$k$ in the range $i+1,i+2,\dots,j-1$. As shown in Ref.~\cite{BBH10}, the spectrum of the
$\su(k|n)$ HS chain (with the correct degeneracies of each level) can be generated through the
formula
\begin{equation}\label{bondstrip}
  \cE(\bsi)=\sum_{i=1}^{N-1}\de(\si_i,\si_{i+1})i(N-i)\,,
\end{equation}
where
\begin{equation}\label{desi}
  \de(\si,\si')=
  \begin{cases}
    1,& \si>\si'\text{ or } \si=\si'\in\{k+1,\dots,k+n\}\\
    0,& \si<\si'\text{ or } \si=\si'\in\{1,\dots,k\}\,,
  \end{cases}
\end{equation}
and the components of the bond vector $\bsi:=(\si_1,\dots,\si_N)$ independently range from $1$ to
$k+n$. In other words, the spectrum of $\cH^{(k|n)}_{\mathrm{HS}}$ is the same as that of a vertex
model with $N+1$ vertices $0,\dots,N$ joined by $N$ bonds with values
$\si_1,\dots,\si_N\in\{1,\dots,k+n\}$, where the energy of the $i$-th bond
is~$\de(\si_i,\si_{i+1})i(N-i)$. It should also be noted that eq.~\eqref{bondstrip} can be used to
compute the spectrum of the $\su(k|n)$ HS chain in subspaces with a well-defined magnon
content~\cite{BBCFG19}, spanned by basis vectors~$\ket{s_1,\dots,s_N}$ all of which contain each
spin value $\al\in\{1,\dots,k+n\}$ a fixed number of times $N_\al$. Indeed, it suffices to
restrict the bond vectors~$\bsi$ in eq.~\eqref{bondstrip} to those containing exactly $N_\al$
times each integer $\al\in\{1,\dots,k+n\}$.

\begin{figure}[t]
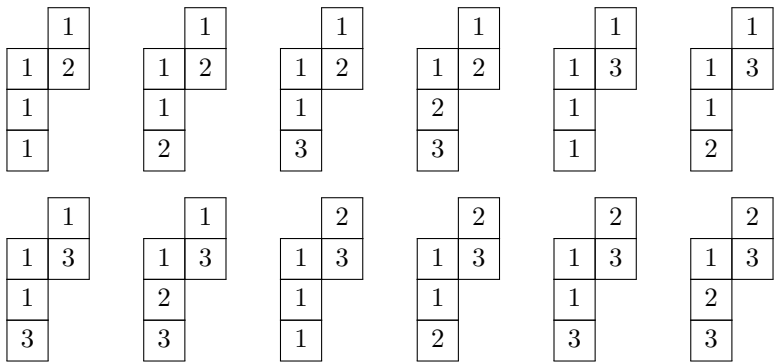

  \hfill\small
\begin{align*}
  &\ytableaushort{\none1,12,1,1
}
\qquad
\ytableaushort{\none1,12,1,2
}
\qquad
\ytableaushort{\none1,12,1,3
}
\qquad
\ytableaushort{\none1,12,2,3
}
\qquad
  \ytableaushort{\none1,13,1,1
}
\qquad
\ytableaushort{\none1,13,1,2
}
\\[2mm]
&\ytableaushort{\none1,13,1,3
}
\qquad
\ytableaushort{\none1,13,2,3
}
\qquad
\ytableaushort{\none2,13,1,1
}
\qquad
\ytableaushort{\none2,13,1,2
}
\qquad
\ytableaushort{\none2,13,1,3
}
\qquad
\ytableaushort{\none2,13,2,3
}
\end{align*}
\hfill
\caption{Allowed $\su(1|2)$ Young tableaux for the motif~$(0,1,0,0)$.
  \label{fig.bs}}
\end{figure}

From eq.~\eqref{bondstrip} it also follows that the numerical values of the energies of the
$\su(k|n)$ HS chain can be expressed by the formula
\begin{equation}\label{motif}
  \cE_{\bde}=\sum_{i=1}^{N-1}\de_i\cdot i(N-i)\,,
\end{equation}
where the vector~$\bde:=(\de_1,\dots,\de_{N-1})\in\{0,1\}^{N-1}$ is a ($\su(k|n)$ supersymmetric)
Haldane motif~\cite{Ha93,HB00,BBHS07}, provided that each motif $\bde$ is assigned an appropriate
degeneracy~$d(\bde)$ (which can be zero). As shown, e.g., in refs.~\cite{KKN97,FG15}, $d(\bde)$ is
equal to the number of different fillings (skew Young tableaux) of the border strip associated
with the motif $\bde$ following a suitable rule. More precisely, given a motif~$\bde$ its
associated border strip is constructed by starting with one box, and then reading the motif from
left to right and adding a box below the $i$-th box provided that $\de_i$ is equal to $0$, or to
its left provided that $\de_i=1$ (cf.~fig.~\ref{fig.bs}). This border strip should then be filled
with the numbers $1,\dots,k+n$ according to the following rules:
\begin{enumerate}[i)]
\item The numbers in each row form a nondecreasing sequence, allowing only the repetition of
  numbers in the range $k+1,\dots,k+n$.
\item The numbers in each column (read from top to bottom) form a nondecreasing sequence,
  allowing only the repetition of numbers in the range $1,\dots,k$.
\end{enumerate}
The equivalence between the descriptions of the spectrum through
eqs.~\eqref{bondstrip}-\eqref{desi} and eq.~\eqref{motif} stems from the fact that to each bond
vector $\bsi$ corresponds in a one-to-one fashion a Young tableau with border strip determined by
the motif $\de_i=\de(\si_i,\si_{i+1})$ and filling given by the components of the vector $\bsi$,
arranged from top to bottom and from right to left. Note also that when $kn\ne0$ (i.e., in the
truly supersymmetric case) the degeneracy of all motifs $\bde$ is obviously nonzero (i.e., all the
sequences of $N-1$ zeros or ones are allowed motifs), while for $kn=0$ this is not the case. More
precisely, in the purely bosonic $(k|0)$ case no sequences with $k$ or more ones are allowed,
while in the purely fermionic one $(0|n)$ no sequences of $n$ or more zeros are allowed. In
general, if $\bde$ is an allowed motif the total degeneracy of the corresponding energy
$\cE_{\bde}$ is given by the sum of the intrinsic degeneracies $d(\bde')$ of all motifs $\bde'$
such that $\cE_{\bde}=\cE_{\bde'}$.

\subsection{Motifs in the even $m$ case}

As remarked in the previous section, the partition function of the $\su(m)$ model~\eqref{Hchain}
with even $m$ factors as the product of the corresponding partition functions of the $\su(1|1)$
and $\su(m/2)$ HS chains~\eqref{suknHS}. It follows that the Hamiltonian $\cH$ of the former model
is isomorphic to the sum of $\cH^{(1|1)}_{\mathrm{HS}}$ and
$\cH^{(m/2)}_{\mathrm{HS}}:=\cH^{(0|m/2)}_{\mathrm{HS}}$ acting on the tensor product of their
respective Hilbert spaces. In other words, there is a unitary transformation
$U:\Si\to(\otimes_{i=1}^{N}\CC^2)\otimes(\otimes_{i=1}^{N}\CC^{m/2})$ such that
\begin{equation}\label{cHcHS}
  U\cH U^{-1}=\cH^{(1|1)}_{\mathrm{HS}}\otimes1+1\otimes\cH^{(m/2)}_{\mathrm{HS}}\,.
\end{equation}
As a consequence, the spectrum of the model~\eqref{Hchain} with even $m$ is obtained by adding up
two arbitrary energies of $\cH^{(1|1)}_{\mathrm{HS}}$ and $\cH^{(m/2)}_{\mathrm{HS}}$, i.e., it
can be expressed by the formula
\begin{equation}\label{cEdds}
  \cE(\bsi,\bsi')=\sum_{i=1}^{N-1}\big[\de(\si_i,\si_{i+1})+\de'(\si'_i,\si'_{i+1}\big]i(N-i)\,,
\end{equation}
where $\bsi=(\si_1,\dots,\si_N)\in\{1,2\}^N$, $\bsi'=(\si'_1,\dots,\si'_N)\in\{1,\dots,m/2\}^N$
are $\su(1|1)$ and $\su(0|m/2)$ bond vectors and $\de(\si_i,\si_{i+1})$
(resp.~$\de'(\si'_i,\si'_{i+1})$) follows the $\su(1|1)$ (respectively $\su(0|m/2)$)
rule~\eqref{bondstrip}. Another important consequence of eq.~\eqref{cHcHS} is the fact that in the
even $m$ case the model is symmetric under a twisted quantum group isomorphic to the direct sum
$Y(\gl(1|1))\oplus Y(\gl(0|m/2))$, due to the symmetry of the $\su(k|n)$ HS chain Hamiltonian
under the Yangian~$Y(\gl(k|n))$~\cite{BBHS07}. As we shall see in the next section, this fact is
reflected in the huge degeneracy of the model's spectrum.

As before, the spectrum of the chain~\eqref{Hchain} can be alternatively obtained through the
formula
\begin{equation}\label{cEdd}
  \cE_{\bde,\bde'}=\sum_{i=1}^{N-1}(\de_i+\de_i')i(N-i)\,,
\end{equation}
where the energies are labeled by a pair of independent motifs $\bde$ and $\bde'$ of respective
types $\su(1|1)$ and $\su(0|m/2)$, each one counted with their respective intrinsic degeneracies
$d(\bde)$ and $d(\bde')$. In fact, it is obvious that the $\su(1|1)$ motif $\bde$ has intrinsic
degeneracy $d(\bde)=2$, since the only two possible fillings of $\bde$ compatible with the
$\su(1|1)$ rule are the bond vectors $(1,\dots, 1,2,1,\dots,1,2,1,\dots,1,\si_{N})$ with
$\si_{N}\in\{1,2\}$ and the $2$'s placed in the positions occupied by the $1$'s in $\bde$. As a
consequence, the number of $\su(1|1)$ motifs is precisely $2^{N-1}$, and the intrinsic degeneracy
of the pair $(\bde,\bde')$ is given by
\begin{equation}\label{degdd}
  d(\bde,\bde')=d(\bde)d(\bde')=2d(\bde')\,.
\end{equation}
\begin{figure}[t]
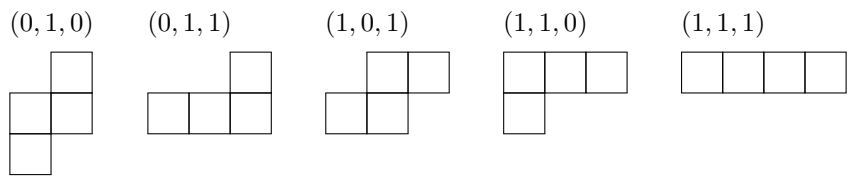

  \hfill\small
  \begin{alignat*}{5}
    &(0,1,0)&\qquad&(0,1,1)&\qquad&(1,0,1)&\qquad&(1,1,0)&\qquad&(1,1,1)\\
    &\ytableaushort{\none{},{}{},{}
    }
    &\qquad&
    \ytableaushort{\none\none{},{}{}{}
    }
    &\qquad&
    \ytableaushort{\none{}{},{}{}\none
    }
    &\qquad&
    \ytableaushort{{}{}{},{}\none\none
    }
    &\qquad&
    \ytableaushort{{}{}{}{}
    }
\end{alignat*}
\hfill
\caption{Allowed $\su(0|2)$ motifs for $N=4$ spins and their corresponding border strips.
  \label{fig.motifs44}}
\end{figure}

Equations~\eqref{cEdd}-\eqref{degdd} provide a concise description of the spectrum of the $\su(m)$
model \eqref{Hchain} with even $m$ which is of great practical value. To illustrate this point,
and exhibit the high degree of degeneracy of the latter model in a concrete example, we shall next
study in some detail the spectrum in the simplest $\su(4)$ case (recall that the $\su(2)$ model is
essentially equivalent to the $\su(1|1)$ HS chain). In this case the motifs $\bde'$ are of
$\su(0|2)$ type, so that they cannot contain two or more consecutive zeros. Let $r\ge0$ be the
number of zeros in~$\bde'$, and denote by $\ka_1+\cdots+\ka_i$ the position of the $i$-th zero. It
is clear that the border strip associated with $\bde'$ consists of $r+1$ horizontal rows of
successive widths (from right to left and top to bottom)
$\ka_1,\dots,\ka_r,\ka_{r+1}:=N-(\ka_1+\cdots+\ka_r)$, and that the only possible fillings of each
of these rows consistent with the $\su(0|2)$ rule are sequences of length $\ka_i$ of the form
$(2,\dots,2,1,\dots,1)$. Both ends of this sequence are fixed for an ``inner'' row (i.e., for the
$i$-th such row with $1<i<r+1$), while the beginning or the end of the sequence are not fixed for
the first and last row, respectively (in particular, if $r=0$ there is only one row whose
corresponding filling can be any sequence of zero or more $2$'s followed by $1$'s). Thus
$d(\bde')=N+1$ for $r=0$ and $d(\bde')=\ka_1\ka_{r+1}\prod_{i=2}^{r}(\ka_i-1)$ for $r>0$; this is
equivalent to Haldane's empirical rule, according to which $d(\bde')$ is obtained by multiplying
out the number of elements in each sequence of consecutive $1$'s in the vector $(1,\bde',1)$. As a
consequence, the degeneracy of a motif pair $(\bde,\bde')$ is given in the $\su(4)$ case by
\begin{equation}\label{degsu4}
  d(\bde,\bde')=
  \begin{cases}
    2(N+1),& r=0\,,\\
    2\ka_1\ka_{r+1}\prod_{i=2}^{r}(\ka_i-1)\,,& r>0\,.
  \end{cases}
\end{equation}

By way of illustration consider, for simplicity, the case $N=4$. There are therefore $2^{N-1}=8$
$\su(1|1)$ motifs, and exactly five allowed $\su(0|2)$ motifs, shown in fig.~\ref{fig.motifs44}
with their corresponding border strips.
We present in table~\ref{tab.en44} the model's energies corresponding to each pair of motifs
$(\bde,\bde')$, indicating their degeneracies by a subscript. Although in principle there could be
as many as $8\times 5=40$ different energy levels, in fact the degeneracy is much higher due to
the extremely simple form of the dispersion relation $\cE(i)$. Indeed, there are only $13$
distinct energy levels, namely
\[
  4_2\,, \quad 6_8\,, \quad 7_{16}\,, \quad 8_2\,, \quad 9_{16}\,, \quad 10_{44}\,, \quad 11_{16}\,,
  \quad 12_8\,, \quad 13_{48}\,, \quad 14_{36}\,, \quad 16_{18}\,, \quad 17_{32}\,, \quad 20_{10}\,,
\]
where the subscripts indicate the degeneracy. Thus the average degeneracy of the levels in this
case is $4^4/13\simeq19.69$.
\begin{table}[t]
\centering%\small
\begin{tabular}{|c|c|c|c|c|c|}
  \hline\small
  $N=m=4$&$(0,1,0)$&$(0,1,1)$&$(1,0,1)$&$(1,1,0)$&$(1,1,1)$\\
\hline 
  $(0,0,0)$& $4_2$& $7_6$& $6_8$& $7_6$& $10_{10}$\\
  $(0,0,1)$& $7_2$&  $10_6$& $9_8$& $10_6$& $13_{10}$\\
  $(0,1,0)$& $8_2$ & $11_6$ & $10_8$ & $11_6$ & $14_{10}$\\
  $(0,1,1)$& $11_2$ & $14_6$ & $13_8$ & $14_6$ & $17_{10}$\\
  $(1,0,0)$& $7_2$ & $10_6$ & $9_8$ & $10_6$ &$ 13_{10}$\\
  $(1,0,1)$& $10_2$ & $13_6$ & $12_8$ & $13_6$ & $16_{10}$\\
  $(1,1,0)$& $11_2$ & $14_6$ & $13_8$ & $14_6$ & $17_{10}$\\
  $(1,1,1)$& $14_2$ & $17_6$ & $16_8$ & $17_6$ & $20_{10}$\\
\hline
\end{tabular}
\caption{\label{tab.en44} Energies of the $\su(4)$ model~\eqref{Hchain} with $N=4$ spins and their
  corresponding degeneracies (indicated by subscripts).}
\end{table}

\subsection{Generalized magnons}

Unlike its HS counterpart~\eqref{HSchainA}, the Hamiltonian~\eqref{Hchain} does not preserve in
general the spin content of a basis state~$\ket{s_1,\dots,s_N}$ due to the presence of spin
reversal operators $S_k$. In other words, the subspaces spanned by basis states with a fixed
number $N_\al$ of spin components $s_i$ equal to $\al$ (for $\al=-M,-M+1\dots,M\equiv\frac{m-1}2$)
are not invariant under the Hamiltonian~\eqref{Hchain}. However, a moment's reflection shows that
the latter Hamiltonian does preserve the larger subspaces
$\Si(N_{\frac12(1-\pi(m))},\dots,N_{M-1},N_M)$ (with $N_{\frac12(1-\pi(m))}+\cdots+N_{M}=N$)
spanned by basis states~$\ket{s_1,\dots,s_N}$ such that
\begin{equation}\label{Si1}
  \big|\big\{i:s_i=\pm\al\big\}\big|=N_\al,\qquad \frac12(1-\pi(m))\le\al\le M\,,
\end{equation}
where $\pi(m)$ is the parity of $m$ and $|A|$ denotes the cardinal of the set~$A$. In other words,
what is conserved is the spin content \emph{in absolute value}. In fact, when $m$ is even $\cH$
preserves the smaller subspaces $\Si_\vep(N_{\frac12},\dots,N_{M-1},N_M)$ defined by~\eqref{Si1}
and the additional condition
\[
  (-1)^{|\{i:s_i>0\}|}=\vep\,,
\]
with $\vep\in\{\pm1\}$. In other words, in this case the \emph{parity} of the number of positive
(or negative) spin components is also conserved\footnote{More generally, in the odd $m$ case $\cH$
  also preserves the subspaces $\Si_\vep(N_0,\dots,N_{M-1},N_M)$ with $N_0=0$.}. We shall
henceforth refer to $(\vep,N_{\frac12},\dots,N_M)$ or $(N_{\frac12(1-\pi(m))},\dots,N_M)$ as the
generalized spin (or magnon) content of the subspace $\Si_\vep(N_{\frac12},\dots,N_{M})$ or
$\Si(N_{\frac12(1-\pi(m))},\dots,N_{M})$, respectively. A natural question at this point is
whether the spectrum of the restriction of the Hamiltonian $\cH$ in eq.~\eqref{Hchain} to the
latter invariant subspaces can be obtained through some kind of restricted motifs. In the even $m$
case,
\[
  \dim\Si_\vep(N_{\frac12},N_{\frac32},\dots,N_{M})=2^{N-1}\frac{N!}{\prod_{\al=1/2}^{M} N_\al!}
\]
is precisely equal to the product of the number of $\su(1|1)$ bond vectors~$\bsi$ with the
restriction
\begin{equation}
  \label{rest11}
  (-1)^{|\{i:\si_i=1\}|}=\vep
\end{equation}
times the number of $\su(0|m/2)$ bond vectors $\bsi'$ such that
\begin{equation}
  \label{restm2}
  \big|\big\{i:\si'_i=\be\big\}\big|=N_{\be-\frac12}\,,\qquad 1\le\be\le\frac m2\,.
\end{equation}
It is thus natural to conjecture that in the even $m$ case the spectrum of $\cH$ on the subspace
with generalized magnon content $(\vep,N_{\frac12},\dots,N_M)$ is given by eq.~\eqref{bondstrip},
provided that the bond vectors~$\bsi$ and~$\bsi'$ are restricted respectively by
eqs.~\eqref{rest11} and~\eqref{restm2}. Remarkably, all our numerical computations fully support
this conjecture.
\begin{table}[t]
\centering%\small
\begin{tabular}{|c|l|}
  \hline\small
  $n$& Spectrum\\
  \hline 
  $0$& $10_1,13_2,14_1,16_1,17_2,20_1$\\
  $1$& $6_1,7_2,9_2,10_6,11_2,12_1,13_6,14_5,16_2,17_4,20_1$\\
  $2$& $4_1,6_2,7_4,8_1,9_4,10_8,11_4,12_2,13_8,14_6,16_3,17_4,20_1$\\
       \hline
\end{tabular}
\caption{\label{tab.magnons} Energy levels of the $\su(4)$ model~\eqref{Hchain} with $N=4$ spins
  in the subspaces $\Si_\pm(n)$ with $n$ spins $\pm1/2$ and their corresponding degeneracies
  (indicated by subscripts).}
\end{table}

For instance, consider the case $N=m=4$ studied above. To begin with, it is clear that in general
the spectrum of $\cH$ on $\Si_\vep(N_{\frac12},\dots,N_M)$ does not depend on $\vep$ (since $\cH$
commutes with $\cT=TS_1$) nor on the order of the $N_\al$'s (indeed, $\cH$ obviously commutes with
the interchange of any pair of spin values $\pm\al$ with another pair $\pm\be$). Thus in the case
under discussion it suffices to check the conjecture on the subspaces~$\Si_+(n):=\Si_+(n,4-n)$
with $n=0,1,2$ spins $\pm1/2$ and an even number of positive spins, of respective dimension $8$,
$32$, and $48$. The allowed $\su(1|1)$ bond vectors are those containing an even number of $1$'s,
namely $\bsi=(1,1,1,1)$, $(2,2,2,2)$ and the six distinct permutations of $(1,1,2,2)$. Likewise,
the allowed $\su(0,2)$ bond vectors are $(2,2,2,2)$ for $n=0$, the four distinct permutations of
$(1,2,2,2)$ for $n=1$, and the six distinct permutations of $(1,1,2,2)$ for $n=2$. We list in
table~\ref{tab.magnons} the energy levels of the Hamiltonian $\cH$ in eq.~\eqref{Hchain} with
$N=m=4$ in each of the subspaces $\Si_\pm(n)$ generated through eq.~\eqref{bondstrip} with the
restrictions~\eqref{rest11}-\eqref{restm2} on the bond vectors~$\bsi$ and $\bsi'$, which exactly
coincides with the result obtained by numerical diagonalization of the Hamiltonian in each of
these subspaces. In particular, we see that the twice degenerate ground state with energy $\cE=4$
belongs to the subspaces~$\Si_\pm(2)$, each of which contains all the distinct energies in the
full spectrum of $\cH$.

\section{Statistical properties of the spectrum}\label{sec.specdeg}

The explicit formulas for the partition function of the chain~\eqref{Hchain} derived in
section~\ref{sec.PF} ---and, for even $m$, the motif-based
formula~\eqref{bondstrip}-\eqref{desi}--- make it possible to exactly compute its spectrum for
relatively high values of $N$ for given $m$. Since, as remarked above, the model with $m=2$ is
essentially the $\su(1|1)$ HS chain, we shall restrict ourselves in this section to the case
$m>2$. More specifically, we shall briefly report several general properties of the spectrum
suggested by our analysis of the $\su(3)$ and $\su(4)$ cases.

To begin with, it is clear from eqs.~\eqref{Zeven} and~\eqref{Zodd} that the partition function of
the model~\eqref{Hchain} has a completely different structure when $m$ is even or odd. In fact, in
the latter case it is not even clear that there is a description of the spectrum in terms of
motifs similar to the one for the even $m$ case presented in the previous section. In spite of
this fact, our results indicate that the spectrum of the model~\eqref{Hchain} has very similar
properties for odd or even $m$.

\begin{figure}[t]
  \includegraphics[width=.49\textwidth]{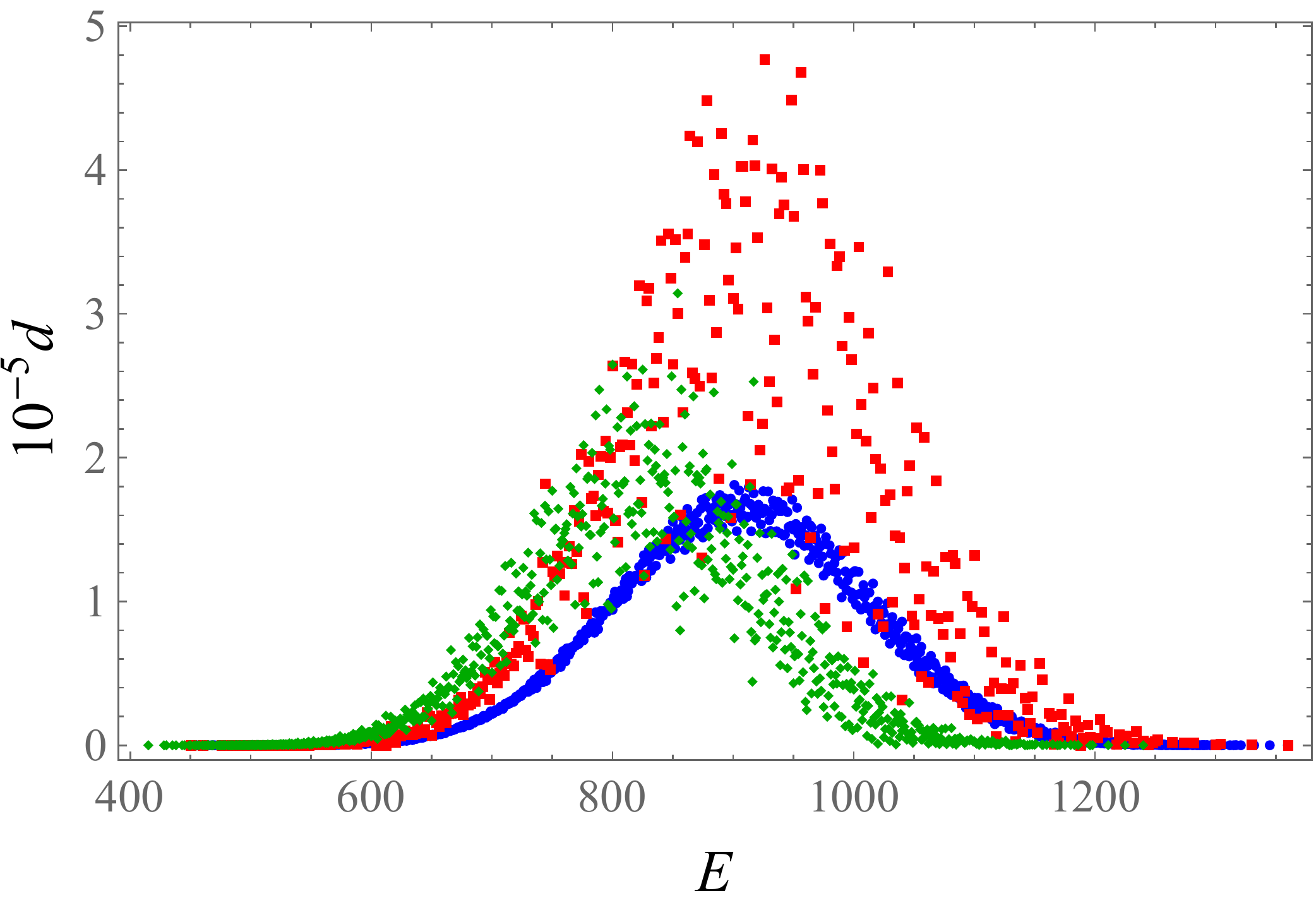}\hfill
  \includegraphics[width=.49\textwidth]{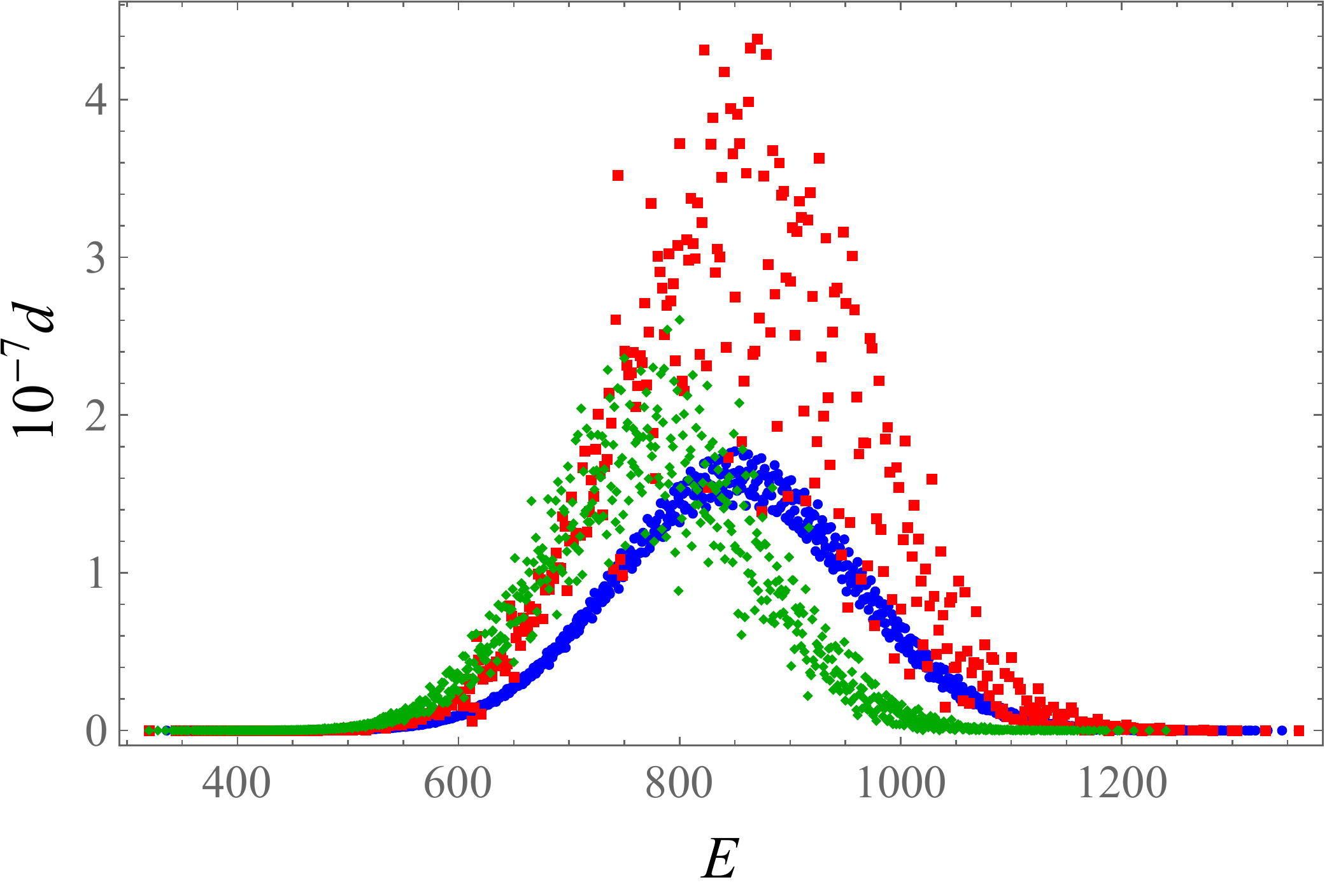}
  \caption{Plot of the degeneracy~$d$ versus the energy~$E$ for the spectra of the $A_{N-1}$ HS
    chain~\eqref{HSchainA} with $J=2$ (red squares), the $D_N$ HS chain~\eqref{HSchainD} with
    $J=1/2$ (green rhombuses) and the model~\eqref{Hchain} (blue circles) for $N=16$ spins and
    $m=3$ (left) or $m=4$ (right).}
  \label{fig.deg}
\end{figure}

\begin{figure}[t]
  \includegraphics[width=.49\textwidth]{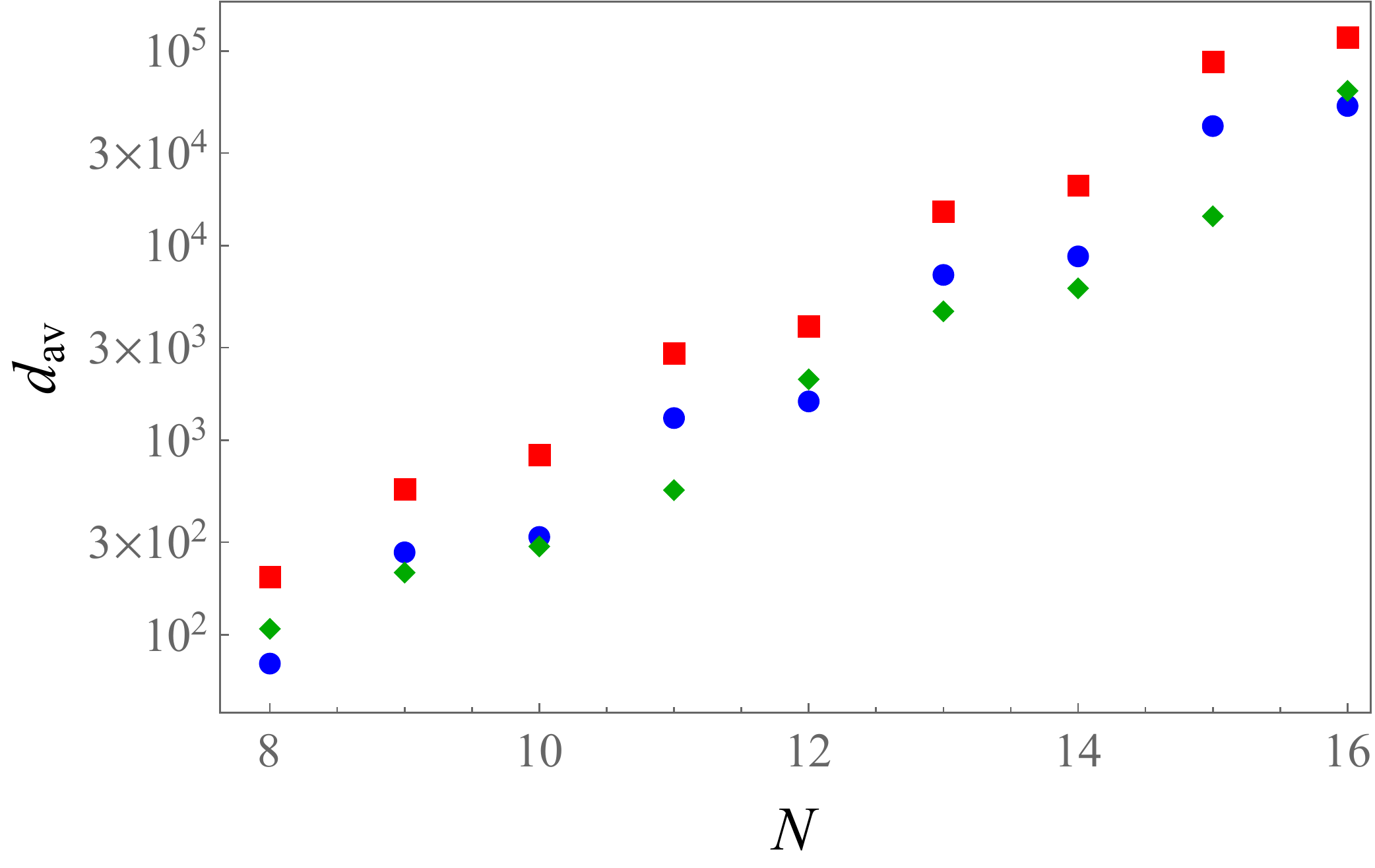}\hfill
  \includegraphics[width=.49\textwidth]{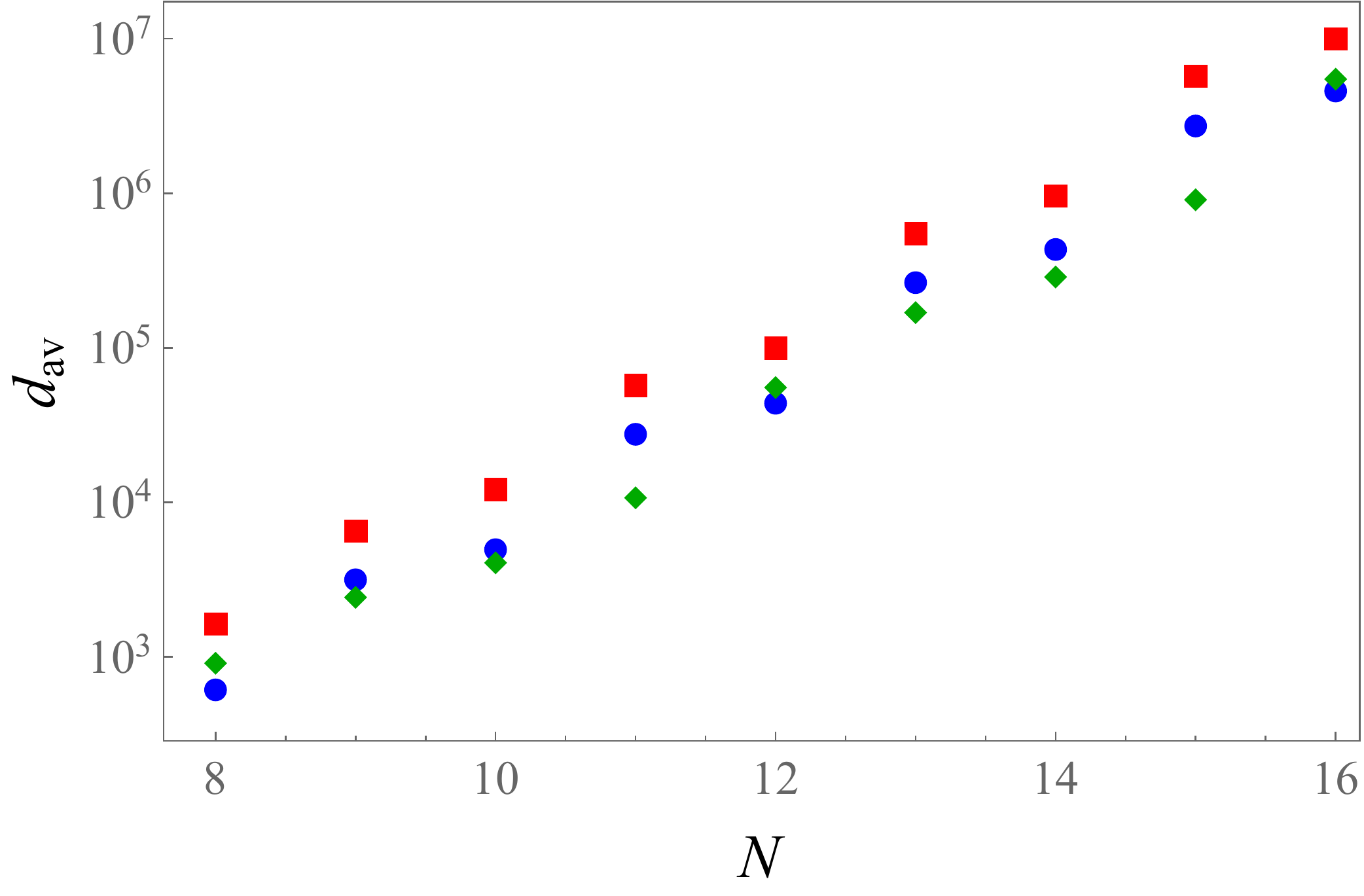}
  \caption{Logarithmic plot of the average degeneracy~$d_{\mathrm{av}}$ of the spectra of the
    model~\eqref{Hchain} (blue dots), the HS chain~\eqref{HSchainA} (red squares) and the $D_N$ HS
    chain~\eqref{HSchainD} (green rhombuses) as a function of the number of spins~$N$ for
    $8\le N\le16$ and $m=3$ (left) or $m=4$ (right).}
  \label{fig.degs}
\end{figure}

More precisely, the first conclusion that we can draw from our evaluation of the partition
function of the chain~\eqref{Hchain} for several values of $N$ is that, regardless of the parity
of $m$, the spectrum exhibits a huge degeneracy. In order to put this statement into perspective,
we have compared the degeneracy of the new chain's spectrum with the degeneracy of the spectra of
the $A_{N-1}$ and $D_N$ HS chains, which is very high due to the Yangian (or twisted Yangian)
invariance of the latter models. For each of these two chains, we have chosen the normalization
parameter $J$ so that their average energy coincides with that of the chain~\eqref{Hchain}. More
precisely, the average energy~$\langle\cH\rangle$ of the latter model can be computed taking into
account that
\begin{equation}\label{trS}
  \tr S_{ij}=\tr \tS_{ij}=m^{N-1},
\end{equation}
so that
\begin{align}
  \langle\cH\rangle:
  &=m^{-N}\tr\cH=\frac14\bigg(1+\frac1m\bigg)\sum_{i<j}\Big(\sin^{-2}(\th_i-\th_j)
    +\cos^{-2}(\th_i-\th_j)\Big)\notag\\
  &=\bigg(1+\frac1m\bigg)\sum_{i<j}\sin^{-2}\left(\tfrac{(i-j)\pi}N\right)
    =\frac 16\bigg(1+\frac1m\bigg)\,N(N^2-1)\,,
    \label{muspec}
\end{align}
where the last sum is evaluated in ref.~\cite{CP78}. The average energy
$\langle\cH_{\mathrm{HS}}\rangle$ of the HS chain can be computed in a similar way, with the known
result~\cite{FG05}
\[
  \big\langle\cH_{\mathrm{HS}}\big\rangle=\frac J{12}\bigg(1+\frac1m\bigg)\,N(N^2-1)\,,
\]
which coincides with $\langle\cH\rangle$ if we take $J=2$. This is consistent with the fact that
the Hamiltonian~\eqref{Hchain} formally reduces to the HS chain Hamiltonian~\eqref{HSchainA} if we
set $S_i=1$ for all~$i$. As to the HS chain of $D_N$ type, proceeding as before and using the
summation formulas for the zeros of Jacobi polynomials in ref.~\cite{ABCOP79} we obtain
\[
  \big\langle\cH_{\mathrm{HS,D}}\big\rangle=\frac J6\,\bigg(1+\frac1m\bigg)\,N(N-1)(2N-1)\,.
\]
Requiring that $\big\langle\cH_{\mathrm{HS,D}}\big\rangle$ agrees with $\langle\cH\rangle$ to
leading order in $N$ we conclude that in this case we should take $J=1/2$. By way of example, in
fig.~\ref{fig.deg} we have plotted the degeneracy of the spectra of the new chain~\eqref{Hchain}
(computed using the exact formula~\eqref{Zeven}-\eqref{Zodd} for its partition function) and of
the HS chains of $A_{N-1}$ and $D_N$ types (with $J=2$ and $J=1/2$, respectively) for $N=16$ spins
and $m=3$ (left) or $m=4$ (right). It is apparent that in both cases the new model's spectrum has
a very high degeneracy, similar to the degeneracy of the spectrum of the $D_N$-type HS chain but
somewhat smaller than that of the $A_{N-1}$ HS chain, even in the odd $m$ case (when the spectrum
cannot be described in terms of motifs).
To make this observation more quantitative, we have computed the average degeneracy of the spectra
of the latter models,
\[
  d_{\mathrm{av}}=\frac{m^N}{n}\,,
\]
where~$n$ denotes the number of distinct energy levels, for $m=3,4$ and $N=8,\dots,16$ spins. The
results, which are presented in fig.~\ref{fig.degs}, again clearly indicate that for sufficiently
high $N$ the spectrum of the new model has a degeneracy similar to that of the $D_N$-type HS chain
and is somewhat less degenerate than that of the original HS chain. It is also apparent from these
plots that the degeneracy of the new chain grows exponentially with $N$, as is typically the case
for Yangian-invariant models~\cite{FG15}. All of these facts strongly suggest that the
model~\eqref{Hchain} possesses some kind of (twisted) Yangian symmetry not only in the even $m$
case, as shown in the previous section, but also for odd $m$. This is typically the case for
similar open chains with long-range interactions like the $BC_N$, $B_N$ and $D_N$ HS chains or the
Simons--Altshuler model~\cite{SA94,BPS95,TS15} and its integrable generalizations~\cite{BFG16}.

\begin{figure}[t]
  \includegraphics[height=.33\textwidth]{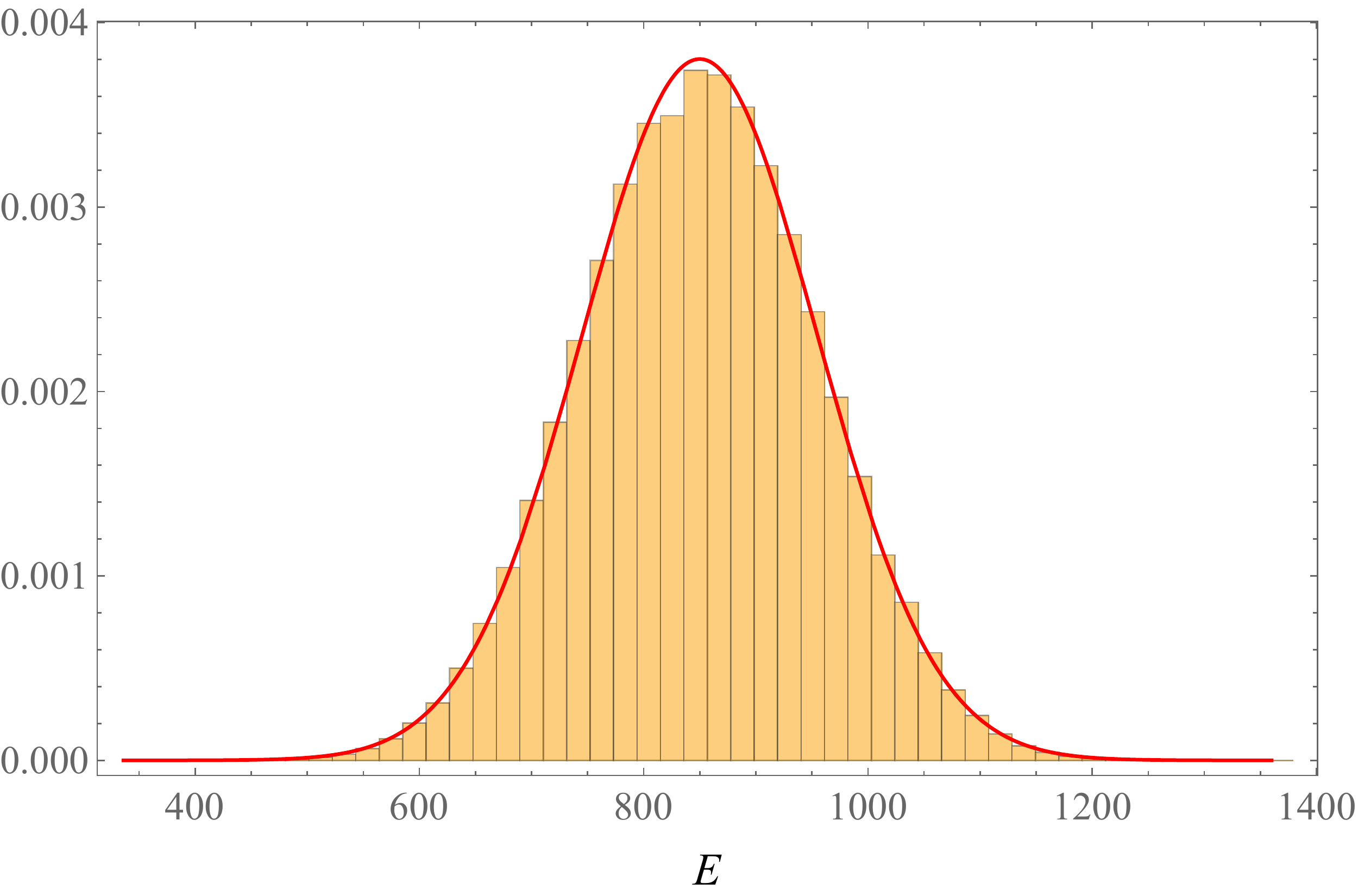}\hfill
  \includegraphics[height=.33\textwidth]{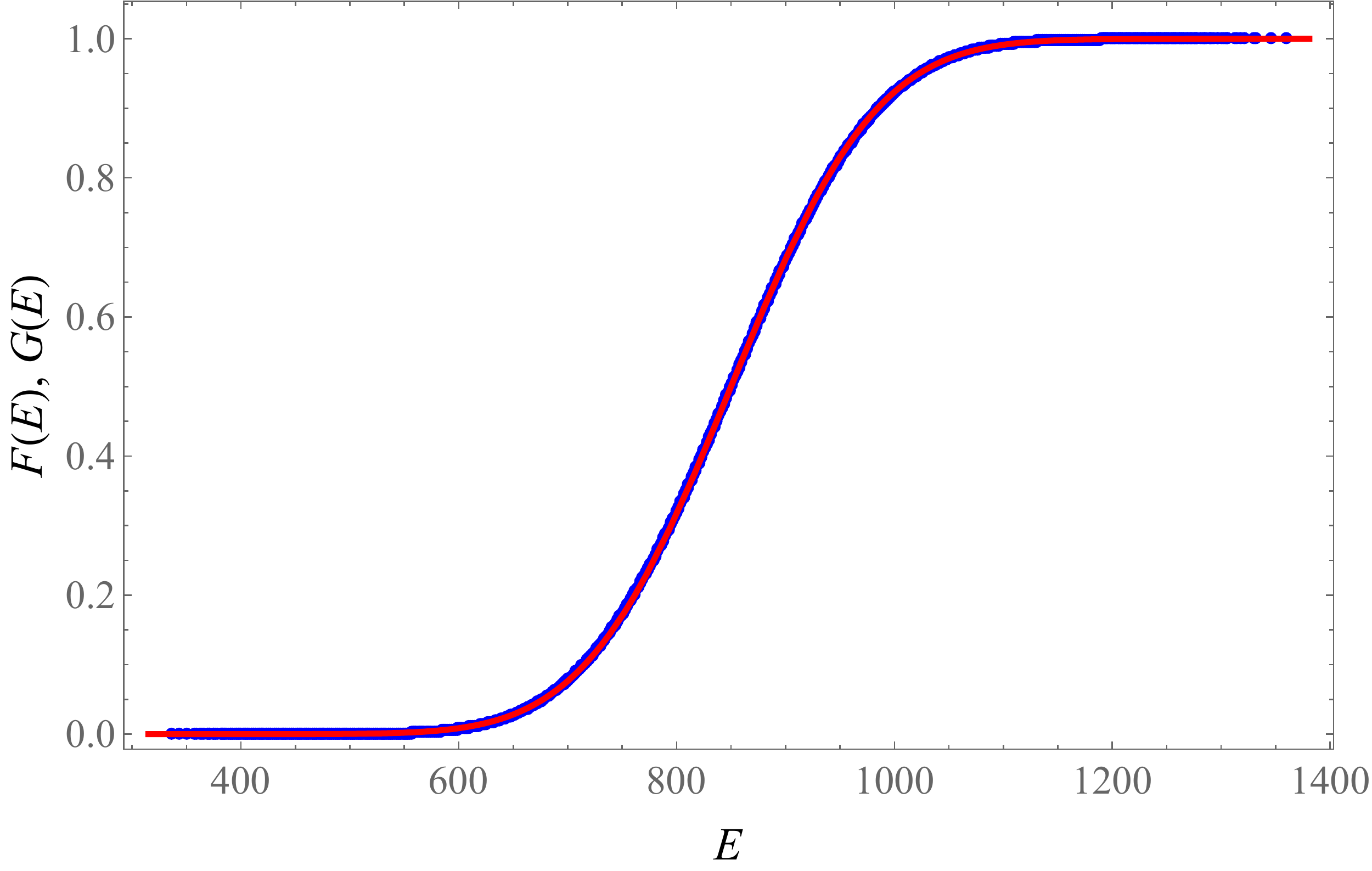}
  \caption{Left: histogram (normalized to unit area) of the degeneracies of the levels of the
    chain~\eqref{Hchain} with $m=4$ and $N=16$ spins compared to the Gaussian distribution (red
    line) with parameters $\mu$ and $\si$ respectively equal to the mean and the standard
    deviation of the model's energy spectrum. Right: plot of the cumulative level density $F(E)$
    of the latter model (blue dots) compared to the cumulative Gaussian distribution~\eqref{Gcum}
    (solid red line).}
  \label{fig.Gauss}
\end{figure}

The plots in fig.~\ref{fig.deg} also seem to indicate that the level density of the new
model~\eqref{Hchain} is Gaussian for sufficiently high $N$, as is the case for all spin chains of
HS type~\cite{EFGR05,EFG10,BFG09,BFG13}. This can be clearly seen for instance from
fig.~\ref{fig.Gauss}, where we have plotted both the histogram of the energy levels and the
cumulative level density of the chain~\eqref{Hchain},
\[
  F(E)=\sum_{i;E_i\le E}d(E_i)\,,
\]
where $d(E_i)$ denotes the degeneracy of the $i$-th level $E_i$, for $N=16$ spins and $m=4$ (the
case $m=3$ is completely similar). It is patent from fig.~\ref{fig.Gauss} that the latter density
is virtually indistinguishable from the cumulative Gaussian distribution
\begin{equation}\label{Gcum}
  G(E)=\frac12\bigg(1+\erf\biggl(\frac{E-\mu}{\sqrt2\si}\biggr)\bigg)
\end{equation}
with parameters $\mu$ and $\si$ respectively equal to the mean~$\langle\cH\rangle$ of the
spectrum, given by eq.~\eqref{muspec}, and its standard
deviation~$(\langle\cH^2\rangle-\langle\cH\rangle^2)^{1/2}$. As shown in appendix~\ref{app.sigma},
the latter quantity can also be evaluated in closed form, with the result
\begin{equation}\label{si2}
  \langle\cH^2\rangle-\langle\cH\rangle^2=\frac1{180}\,\bigg(1-\frac1{m^2}\bigg)N(N^2-1)(2N^2+7)
  -\frac1{12m^2}\,N(N^2-1)(1-\pi(m))\,.
\end{equation}
Note that~\eqref{si2} differs from its counterpart for the HS chain with $J=2$ computed in
ref.~\cite{FG05}, although both quantities do coincide at leading order $O(N^5)$.

\section{Conclusions and outlook}\label{sec.concout}

In this paper we have introduced a new class of open, translationally invariant spin chains with
long-range interactions, which includes the original Haldane--Shastry chain as a particular
(degenerate) case. The Hamiltonian of the new models, which contains both spin permutation and
(polarized) spin reversal operators, turns out to be invariant under ``twisted'' translations
combining an ordinary translation with a spin flip at one end of the chain. This remarkable
invariance is one of the key properties shared by all members of the new class, regardless of the
form of the spin-spin interactions. In fact, the models of this type are fundamentally different
from all spin chains of Haldane--Shastry type studied so far, since their Hamiltonian cannot be
obtained in the usual way from an extended root system. We have constructed a new elliptic chain
of this class which smoothly interpolates between the twisted XXX Heisenberg model and what is
perhaps the simplest model of the new type, in which the spin-spin exchange interaction is
proportional to the inverse square of the (chord) distance. We have computed in closed form the
partition function of the latter model applying Polychronakos's freezing trick to a related spin
dynamical model, whose spectrum we have also determined. When the number $m$ of internal degrees
of freedom is even, the formula for the partition function implies that the new model is
isomorphic to the sum of an $\su(1|1)$ and an $\su(m/2)$ (antiferromagnetic) HS chain Hamiltonian
acting on the tensor product of their respective Hilbert spaces (cf.~eq.~\eqref{cHcHS}). In
particular, this implies that the even $m$ model is symmetric under the direct sum of the Yangians
$Y(\gl(1|1))$ and $Y(\gl(0|m/2))$. The latter property has also been used to derive a simple
description of the spectrum in the even $m$ case in terms of a pair of independent Haldane motifs
of types $\su(1|1)$ and $\su(0|m/2)$. With the help of the partition function, we have also
analyzed the spectrum of the new chain for several values of $m$ and $N$, studying some of its
main statistical properties. Our analysis clearly suggests that the level density is Gaussian for
sufficiently large $N$, as is the case with spin chains of HS type. Moreover, our results indicate
that the model's spectrum has a huge degeneracy, comparable to that of well-known Yangian- or
twisted Yangian-invariant models like the HS chain and its open versions. This is a strong
indication that the new chain possesses an underlying (twisted) Yangian symmetry also in the odd
$m$ case.

Our results open up several lines for future research. To begin with, the explicit description of
the spectrum for even $m$ in terms of motifs makes it possible to study the thermodynamics of the
model through the inhomogeneous transfer matrix approach successfully applied to similar
chains~\cite{EFG12,FGLR18,BBCFG19b}. Finding an explicit realization of the generators of the
Yangian symmetry group would also be highly desirable in this case. Another natural line of
inquiry in this respect is to determine whether there is a similar motif-based description of the
spectrum in the odd $m$ case. A related challenging and relevant open problem is to establish the
existence of some kind of Yangian symmetry and its explicit form in the odd $m$ case which, as
mentioned above, is suggested by the huge degeneracy of the spectrum, and thus prove the model's
integrability. It would also be of interest to find other solvable and/or integrable members of
the new class introduced in this paper. For instance, the version with arbitrarily polarized spin
reversal operators of the chain studied here, as well as its supersymmetric counterpart, should
also be solvable with the method developed in this paper. Another problem worthy of investigation
is the study of the (possibly partial) solvability of models with other spin-spin interactions,
like the elliptic chain introduced in section~\ref{sec.spindm}.

As mentioned in the Introduction, the Hamiltonians of several spin chains with long-range
interactions, including the original HS chain, are the parent Hamiltonians of infinite MPSs states
(usually the ground state) constructed from certain rational CFTs like the $\su(m)_1$ WZNW model.
Although the models introduced in this work share many properties with the latter spin chains, a
crucial difference between both types of models is the presence of spin reversal operators in the
Hamiltonian of the new ones. It would therefore be worth investigating whether the new models can
also be constructed, at least in some cases, from appropriate CFTs, and if so what kind of CFTs
would appear in this context. On a more speculative note, it could also be of interest to explore
whether any of the new models are relevant in connection with the AdS-CFT conjecture, as has
proved to be the case with other translationally invariant spin chains like the Inozemtsev chain.

\appendix

\section{Auxiliary Hamiltonian $H'$ and Dunkl operators}\label{app.specHaux}
In this appendix we introduce an auxiliary scalar operator $H'$ that shall be used in
appendix~\ref{app.specpiper} to compute the spectrum of the dynamical Hamiltonian $H$ in
eq.~\eqref{Hspin}. A key property of this operator is the fact that it can be expressed as the sum
of the squares of a certain family of commuting Dunkl operators, explicitly defined in the second
part of this appendix.

\begin{figure}[t]
  \includegraphics[height=.5\textwidth]{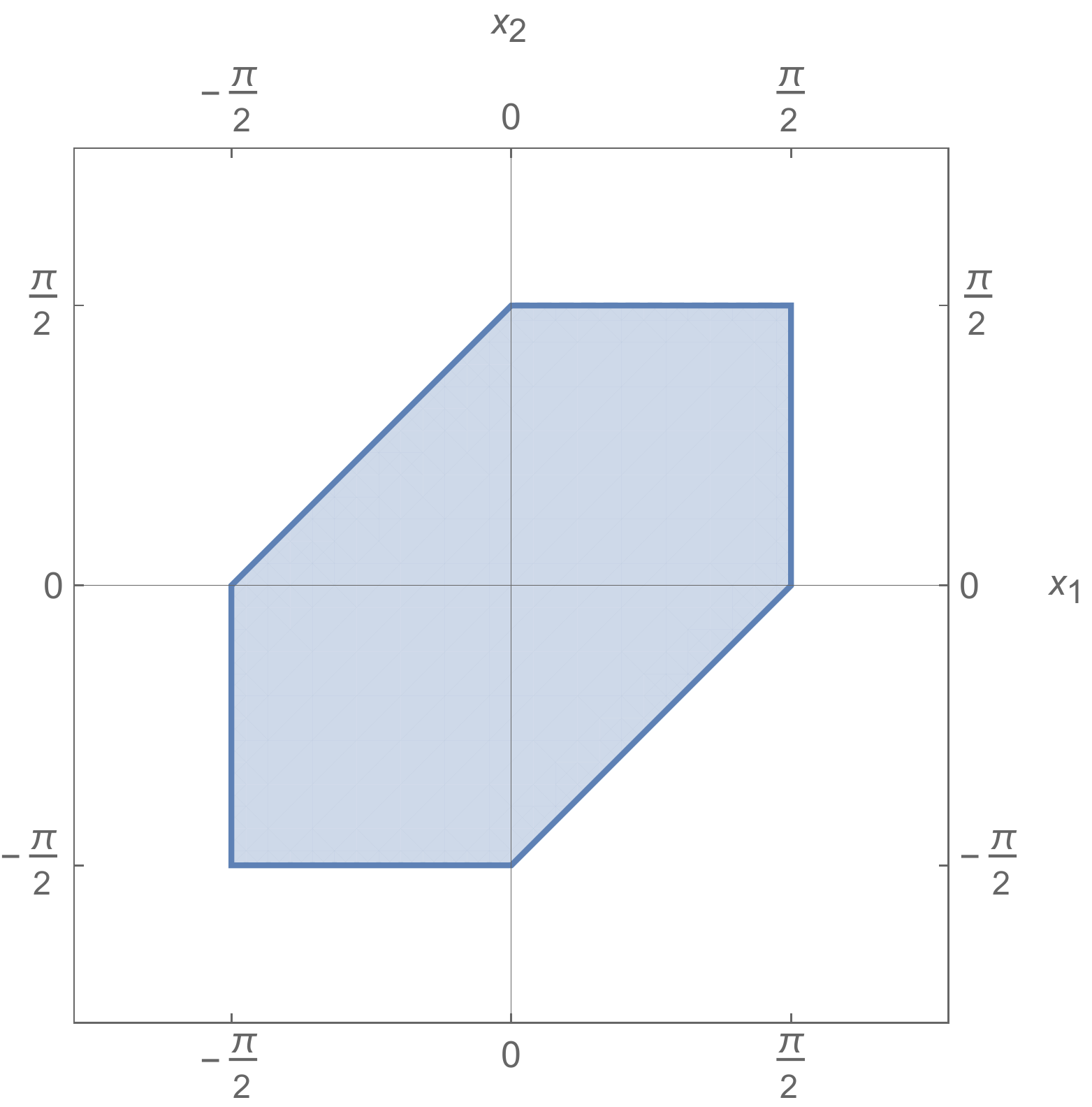} \hfill \includegraphics[height=.5\textwidth]{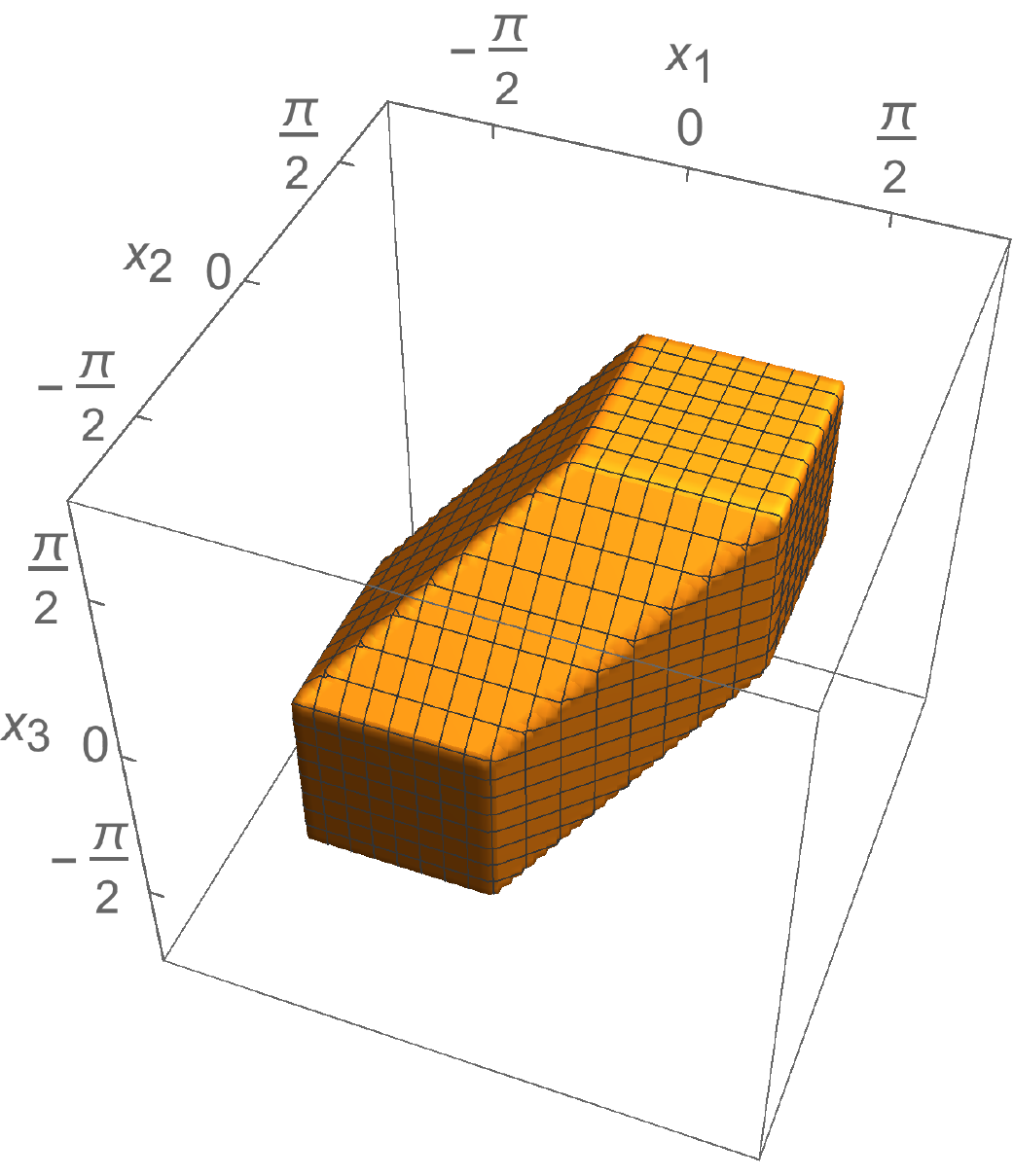}
  \caption{Configuration space~$A'\cap[-\pi/2,\pi/2]^N$ of the auxiliary operator $H'$ for $N=2$
    (left) and $N=3$ (right). In general, the configuration space is an irregular $N$-polytope
    with $N(N+1)$ faces.}
  \label{fig.conf}
\end{figure}

More precisely, let us define
\begin{equation}\label{Hp}
  H'=-\De+2a\sum_{i<j}\bigg[\frac{a-K_{ij}}{\sin^2(x_i-x_j)}
  +\frac{a-\tK_{ij}}{\cos^2(x_i-x_j)}\bigg]\,,
\end{equation}
where $K_{ij}$ denotes the coordinate permutation
operator acting on a scalar function~$f(\bx)$ as
\begin{equation}\label{Kij}
  (K_{ij}f)(x_1,\dots,x_i,\dots,x_j,\dots,x_N)=f(x_1,\dots,x_j,\dots,x_i,\dots x_N)\,,
\end{equation}
$\tK_{ij}:=K_{ij}K_iK_j$, and $K_l$ is the translation operator defined by
\begin{equation}\label{Ki}
  (K_lf)(x_1,\dots,x_l,\dots,x_N)=f(x_1,\dots,x_l+\pi/2,\dots,x_N)\,.
\end{equation}
Note that the operator~$K_l$ is obviously related to the star operator defined in
section~\ref{sec.model} by $z_l^*=\e^{2\iu (K_l x_l)}$. Before proceeding further, a comment on
the domain of the operator~$H'$ is in order. Indeed, the presence of the operators $K_{ij}$ and
$K_iK_j$ in $H'$ entails that we have to enlarge its natural configuration space~\eqref{Adef} to a
region invariant under the action of the latter operators. To begin with, the invariance under the
permutation operators $K_{ij}$ leads to the region
\[
  A'=\{\bx\in\RR^N:|x_i-x_j|<\pi/2,\en 1\le i<j\le N\}.
\]
On the other hand, for the operators $K_{ij}$ and $K_iK_j$ to generate the Weyl group of $D_N$
type, it is necessary that $(K_iK_j)^2=1$ for all~$i\ne j$. In fact, as will become clear in the
sequel, the stronger condition~$K_i^2=1$ for all~$i$ is actually needed in what follows. In view
of this condition it is intuitively clear that we must take as configuration space of $H'$ the
region $A'$, identifying $\bx$ with $\bx+\pi\bev_i$ in the definition of $K_i$\footnote{It is
  straightforward to check that $A'$ is invariant under the operators $K_iK_j$ if we take into
  account the identifications $\bx\equiv\bx+\pi\bev_i$ for all $i$.}. In a more technical
language, we shall take as domain of $H'$ an appropriate dense subspace~$\cP$ of the Hilbert space
\[
  \fH'=L^2([-\pi/2,\pi/2]^N\cap A')
\]
on which $K_i^2=1$ for all $i$ (see figure~\ref{fig.conf} for a sketch of the configuration space
$[-\pi/2,\pi/2]^N\cap A'$ in two and three dimensions). The space~$\cP$ will be chosen as the span
of the $\pi$-periodic trigonometric functions\footnote{We shall see below that the functions
  $\vp_\bn$ actually belong to the domain of $H'$, i.e., that $H'\vp_\bn$ is regular on the
  singular hyperplanes $x_i=x_j$ and $|x_i-x_j|=\pi/2$ with $i\ne j$.}
\begin{equation}
  \label{phi}
  \vp_{\bn}(\bx)=\mu(\bx)\e^{2\iu\bn\cdot\bx}\,,\qquad\bn:=(n_1,\dots,n_N)\in\ZZ^N\,,
\end{equation}
where the factor
\begin{equation}\label{mu}
  \mu=\prod_{í<j}\big|\sin\bigl(2(x_i-x_j)\bigr)\big|^a
\end{equation}
(which, as we shall see, is the ground state of~$H'$) is included for later convenience. Indeed,
the span of the functions~\eqref{phi} is dense in $L^2([-\pi/2,\pi/2]^N)$, and hence in $\fH'$. It
is also important to realize that the conditions~$K_i^2=1$ for all $i$ on $\cP$ entail the
quantization of the system's total momentum, which causes the spectrum of $H'$ on $\fH'$ to be
discrete.

We next introduce the differential-difference Dunkl operators~\cite{Du98,FGGRZ01b}
\begin{equation}
  \label{Dunkl}
  J_l=\iu\,\frac\pd{\pd x_l}-a\bigg\{\iu\sum_{\mathclap{j;j\ne l}}\big[\cot(x_l-x_j)K_{lj}
  -\tan(x_l-x_j)\tK_{lj}\big]+\sum_{j=1}^{l-1}(K_{lj}+\tK_{lj})-\sum_{\mathclap{j=l+1}}^N(K_{lj}+\tK_{lj})
  \bigg\},
\end{equation}
where~$l=1,\dots,N$ and $K_{lj}=K_{jl}$ for~$l>j$. The operators~$J_l$ are related to the
analogous operators $J_l^0(\bz)$ in ref.~\cite{FGGRZ01b} by
\begin{equation}\label{JJ0}
  J_l=-2\mu\,
  J_l^0(\bz)\mu^{-1}-a\sum_{j=1}^{l-1}(K_{lj}+\tK_{lj})+a\sum_{\mathclap{j=l+1}}^N(K_{lj}+\tK_{lj}),
\end{equation}
where $\bz:=(z_1,\dots,z_N)$ with $z_k=\e^{2\,\iu x_k}$ and the inessential parameter $m$ in the
latter reference has been set to~$0$. Note also that the action of the operator~$K_i$ on the
variable~$z_i$ is simply $K_iz_i=z_i^*=-z_i$, akin to the action of $S_i$ on the spin
variable~$s_i$. This observation, in fact, lends additional motivation to the definition of the
action of $K_i$ on the physical coordinate $x_i$ as a translation by~$\pi/2$.

A long but standard explicit computation shows that the operators~$J_i$ ---unlike their
counterparts~$J_l^0(\bz)$--- form a commuting family, i.e.,
\[
  [J_i,J_j]=0\,,\qquad\all i,j=1,\dots,N\,.
\]
It should be noted that for~$N\ge3$ it is essential for the validity of this result that the
operators $K_i$ satisfy $K_i^2=1$, as opposed to the weaker condition~$(K_iK_j)^2=1$ for
all~$i\ne j$ stemming from the~$D_N$ Weyl group algebra. Moreover, a similar calculation proves
that the auxiliary operator~$H'$ can be expressed in terms of the Dunkl operators~$J_i$ as
\begin{equation}
  \label{HpJs}
  H'=\sum_iJ_i^2\,.
\end{equation}
Again, for~$N\ge3$ the above identity requires that $K_i^2=1$ for all $i$. In
appendix~\ref{app.specpiper} we shall rely on eq.~\eqref{HpJs} to explicitly compute the spectrum
of~$H'$ on $\fH'$. This result will then be used in section~\ref{sec.PF} to derive the spectrum of
the spin dynamical model~$H$ in the CM frame.

\section{Spectrum of the dynamical models $H'$ and $H|_{\fH}$}\label{app.specpiper}

In this appendix we shall first compute the spectrum of the auxiliary Hamiltonian $H'$ in its
natural Hilbert space $\fH'$. We shall then use this result to derive the spectrum of the
Hamiltonian $H$ in a suitable subspace of $\fH\subset\fH'\otimes\Si$, in which the action of the
latter operator coincides with that of $H'$.

\subsection{Spectrum of $H'$}

The starting point in the computation of the spectrum of~$H'$ on $\fH'$ is the fact that the
operators~$J_l$ leave invariant the (infinite-dimensional) subspace~$\cP_n\subset\cP$ spanned by
the functions $\vp_{\bn}(\bx)$ with $|\bn|:=\sum_in_i=n$, for an arbitrary (possibly negative)
integer~$n$. In fact, the operators $J_l$ also admit the finite-dimensional invariant
subspaces~$\cR_p$ ($p=0,1,\dots$) spanned by the latter monomials with multiindices satisfying the
conditions $|n_i|\le p$ for $i=1,\dots,N$. By eq.~\eqref{HpJs}, the auxiliary operator~$H'$ will
also leave invariant the subspaces $\cP_n$ and $\cR_p$, and hence their intersection. We shall
construct a non-orthonormal (i.e., Schauder) basis~$\cB$ of~$\fH'$ by appropriately ordering the
set~$\{\vp_\bn(\bx):\bn\in\ZZ^N\}$, and then show that the operator $H'$ is upper triangular with
respect to this basis. Finally, we shall derive the spectrum of~$H'$ by explicitly computing its
diagonal matrix elements in the basis~$\cB$.

In order to simplify the calculation, we shall apply the change of variables $z_k=\e^{2\,\iu x_k}$
and the pseudo-gauge transformation with gauge factor~$\mu$ to the Dunkl operators~$J_l$, and thus
work with the operators
\[
  J_i^*:= \mu^{-1}J_i\,\mu = -2z_i\frac{\pd}{\pd z_i}-a\sum_{j;j\ne i}\bigg[
  \frac{z_i+z_j}{z_i-z_j}\,(1-K_{ij}) +
  \frac{z_i-z_j}{z_i+z_j}\,(1-\tK_{ij})+\sgn(i-j)(K_{ij}+\tK_{ij}) \bigg],
\]
where~$\sgn(x)$ is the sign of~$x$. Defining
\begin{equation}\label{phin}
  \phi_\bn(\bz)=\mu(\bx)^{-1}\vp_\bn(\bx)=\prod_kz_k^{n_k},
\end{equation}
where $n_k\in\ZZ$, we then have
\[
  J_i\vp_{\bn}=\mu J_i^*\phi_\bn\,.
\]
It shall also be necessary for what follows to define a partial order in the set of
monomials~$\phi_\bn$. This is done in four stages, as we shall now explain. To begin with, we
order the multiindices~$\bn\in\ZZ^N$ by $p(\bn):=\max\{|n_i|:i=1,\dots,N\}$, and then (for
$p(\bn)=p(\bn')$) by $n:=\sum_{i}n_i$. We next define~$[\bn]\in\ZZ^N$ as the rearrangement of the
multiindex~$\bn\in\ZZ^N$ whose components are weakly decreasing, i.e., such
that~$[\bn]_1\ge[\bn]_2\ge\cdots\ge[\bn]_N$. The multiindices with given $p(\bn)$ and $n$ are then
partially ordered using their rearrangements~$[\bn]$, i.e., defining that $\bn$ precedes $\bn'$ if
the first nonzero difference~$[\bn]_i-[\bn']_i$ is negative. The above prescription clearly
defines a total order among weakly decreasing multiindices and a \emph{partial} order in $\ZZ^N$
which we shall denote by $\prec\,$, with $0\equiv(0,\dots,0)$ as first element. We shall then say
that $\phi_\bn\prec\phi_{\bn'}$ if and only if~$\bn\prec\bn'$. We shall also use the
notation~$\bn\preceq\bn'$ to indicate that~$\bn'\not\prec\bn$ (i.e., either $\bn\prec\bn'$ or
$[\bn]=[\bn']$), and similarly for $\phi_\bn\preceq\phi_{\bn'}$. It is obvious from the previous
definition that the partial order thus defined is invariant under coordinate permutations, i.e.,
\[
  \phi_\bn\prec\phi_{\bn'}\en\implies\en W\phi_\bn\prec W\phi_{\bn'}
\]
where $W$ is an arbitrary element of the permutation group generated by the operators~$K_{ij}$. Of
course, the partial order~$\prec$ can be equivalently defined on the functions~$\vp_\bn$ by
setting~$\vp_\bn\prec\vp_{\bn'}$ if and only if~$\bn\prec\bn'$. Note, finally, that~$\prec$ can be
promoted to a total order by setting the precedence of two multiindices which differ by a
permutation of their components in an arbitrary way. For instance, if we choose lexicographic
order then $(-1,\dots,-1)$ is the second element (this is always the case due to the partial
order), $(-1,\dots,-1,0)$ is the third, $(-1,\dots,-1,0,-1)$ the fourth, etc.

We start by computing~$J_i^*\phi_\bn$ on a monomial~$\phi_\bn$. To this end, note first of all
that
\begin{align*}
  \frac{z_i+z_j}{z_i-z_j}\,(1-K_{ij})z_i^{n_i}z_j^{n_j}
  &=(z_i+z_j)\,
    \frac{z_i^{n_i}z_j^{n_j}-z_i^{n_j}z_j^{n_i}}{z_i-z_j}\\
  &=\sgn(n_i-n_j)(z_iz_j)^{\min(n_i,n_j)}(z_i+z_j)\,\frac{z_i^{|n_i-n_j|}-z_j^{|n_i-n_j|}}{z_i-z_j},
\end{align*}
where by definition~$\sgn(0)=0$. Since the exponents in the numerator are nonnegative we can carry
out the division, thus easily obtaining
\begin{multline} \label{Kijzizj} \frac{z_i+z_j}{z_i-z_j}\,(1-K_{ij})z_i^{n_i}z_j^{n_j}
  =\sgn(n_i-n_j)\Big(z_i^{n_i}z_j^{n_j}+z_i^{n_j}z_j^{n_i}\\+2\sum_{k=1}^{|n_i-n_j|-1}
  z_i^{\max(n_i,n_j)-k}z_j^{\min(n_i,n_j)+k}\Big).
\end{multline}
Similarly,
\begin{multline}\label{tKijzizj}
  \frac{z_i-z_j}{z_i+z_j}\,(1-\tK_{ij})z_i^{n_i}z_j^{n_j}
  =(-1)^{n_j}\sgn(n_i-n_j)\Big(z_i^{n_i}(-z_j)^{n_j}+z_i^{n_j}(-z_j)^{n_i}\\
  +2\sum_{k=1}^{|n_i-n_j|-1} z_i^{\max(n_i,n_j)-k}(-z_j)^{\min(n_i,n_j)+k}\Big),
\end{multline}
and
\begin{equation}\label{KijtKij}
  (K_{ij}+\tK_{ij})z_i^{n_i}z_j^{n_j}=z_i^{n_j}z_j^{n_i}\big(1+(-1)^{n_i+n_j}\big).
\end{equation}
Note, in particular, that the right-hand side of eqs.~\eqref{Kijzizj}--\eqref{KijtKij} obviously
belongs to~$\cP_n^*:=\mu^{-1}\cP_n$ with $n=|\bn|$, i.e., the space spanned by
monomials~$\phi_{\bn}$ with integer exponents~$n_k$ ($k=1,\dots,N$) of total degree~$|\bn|=n$.
Hence~$J_i^*\cP_n^*\subset\cP_n^*$, or equivalently $J_i\cP_n\subset\cP_n$, as we had previously
claimed. In the same way it is established that~$J_i^*\cR_p^*\subset\cR_p^*:=\mu^{-1}\cR_p$, and
hence $J_i\cR_p\subset\cR_p$. It is also easy to convince oneself that all the terms in the sums
in eqs.~\eqref{Kijzizj}-\eqref{tKijzizj} precede the monomial~$z_i^{n_i}z_j^{n_j}$ with respect to
the partial order defined above, since the exponents of $z_i$ and $z_j$ in each of them range over
$ \min(n_i,n_j)+1$ and $\max(n_i,n_j)-1$ and their sum equals $n_i+n_j$. This is easily seen to
imply that
\begin{align}
  \frac{z_i+z_j}{z_i-z_j}\,(1-K_{ij})\phi_{\bn}
  &=\sgn(n_i-n_j)(z_i^{n_i}z_j^{n_j}+z_i^{n_j}z_j^{n_i})\prod_{k;k\ne
    i,j}z_k^{n_k}+\text{l.o.t.}\,,
    \label{Kijact}\\  
  \frac{z_i-z_j}{z_i+z_j}\,(1-\tK_{ij})\phi_\bn
  &=\sgn(n_i-n_j)\big(z_i^{n_i}z_j^{n_j}+(-1)^{n_i+n_j}z_i^{n_j}z_j^{n_i}\big)\prod_{k;k\ne
    i,j}z_k^{n_k}+\text{l.o.t.}\,,
    \label{tKijact}
\end{align}
where~l.o.t. stands for a linear combination of monomials preceding~$\phi_\bn$. From
eqs.~\eqref{KijtKij}--\eqref{tKijact} it immediately follows that~$\phi_\bn$ does not precede any
monomial in~$J_i^*\phi_{\bn}$, i.e., $J_i^*\phi_\bn$ is of the form
\begin{equation}
  \label{Jiphinnp}
  J_i^*\phi_\bn=\sum_{\bn'\preceq\bn}c_{\bn'\bn}\phi_{\bn'}
\end{equation}
for appropriate coefficients~$c_{\bn'\bn}\in\CC^N$.

The last observation can be considerably strengthened if we assume that~$\bn\in\big[\ZZ^N\big]$,
i.e., if $n_1\ge n_2\ge\cdots\ge n_N$. Indeed, in this case
\[
  \sgn(n_i-n_j)=(1-\de_{n_in_j})\sgn(j-i),
\]
so that the coefficient of $z_i^{n_j}z_j^{n_i}$ in
\[
  \bigg(\frac{z_i+z_j}{z_i-z_j}\,(1-K_{ij})+\frac{z_i-z_j}{z_i+z_j}\,(1-\tK_{ij})
  +\sgn(i-j)(K_{ij}+\tK_{ij})\bigg)z_i^{n_i}z_j^{n_j}
\]
equals
\[
  \big[\sgn(j-i)(1-\de_{n_in_j})+\sgn(i-j)\big]\big(1+(-1)^{n_i+n_j}\big) =2\de_{n_in_j}\sgn(i-j)
\]
and therefore
\begin{multline*}
  \bigg(\frac{z_i+z_j}{z_i-z_j}\,(1-K_{ij})+\frac{z_i-z_j}{z_i+z_j}\,(1-\tK_{ij})
  +\sgn(i-j)(K_{ij}+\tK_{ij})\bigg)z_i^{n_i}z_j^{n_j}\\=2\sgn(j-i)(1-2\de_{n_in_j})z_i^{n_i}z_j^{n_j}
  +\text{l.o.t.}
\end{multline*}
Multiplying both sides of the latter equation by~$\prod_{k;k\ne i,j}z_k^{n_k}$ we finally obtain
\begin{equation}
  \label{Jiphin}
  J_i^*\phi_\bn=\la_i(\bn)\phi_\bn+\text{l.o.t.},
\end{equation}
where~l.o.t.~consists of monomials $\phi_{\bn'}\prec\phi_\bn$ and
\begin{align*}
  \la_i(\bn)&=-2n_i+2a\sum_{j;j\ne i}\sgn(i-j)(1-2\de_{n_i,n_j})\\
            &=
              -2n_i+2a(2i-N-1)+4a\big|\{j:j>i,\ n_i=n_j\}\big|
              -4a\big|\{j:j<i,\ n_i=n_j\}\big|,
\end{align*}
with~$|A|$ denoting the cardinal of a set~$A$. The above formula can be slightly simplified by
introducing the notation
\[
  l(n_i)=\min\{j:n_j=n_i\}\,,\qquad \#(n_i)=|\{j:n_j=n_i\}|,
\]
in terms of which
\begin{equation}
  \label{laibn}
  \la_i(\bn)=-2n_i+2a\big[4l(n_i)+2\#(n_i)-2i-N-3\big]\,.
\end{equation}

We shall next make use of eqs.~\eqref{Jiphinnp}-\eqref{laibn} above to obtain the spectrum
of~$H'$. More precisely, we shall see that the action of~$H'$ is upper triangular on the
(non-orthonormal) basis $\cB=\{\vp_\bn:\bn\in\ZZ\}$ of~$\fH'$, ordered with any ordering
compatible with $\prec$, and shall explicitly compute its diagonal matrix elements.

We start by evaluating $H'\vp_\bn$ when the multiindex~$\bn$ belongs to $[\ZZ^N]$, i.e., is
non-increasing. To this end, note first of all that
\[
  H'\vp_\bn=\mu\cdot (\mu^{-1}H'\mu)\phi_\bn=\mu\sum_i(J_i^*)^2\phi_\bn\,.
\]
Since by assumption~$\bn\in[\ZZ^N]$, by eq.~\eqref{Jiphin} we have
\[
  (J_i^*)^2\phi_\bn=\la_i(\bn)J_i^*\phi_n+J_i^*(\text{l.o.t})
  =\la_i(\bn)J_i^*\phi_\bn+\text{l.o.t.}=\la_i(\bn)^2\phi_{\bn}+\text{l.o.t.}\,,
\]
where in the second equality we have used eq.~\eqref{Jiphinnp}. Hence in this case
\begin{equation}\label{Hpvpn}
  H'\vp_\bn=\Big(\sum_i\la_i(\bn)^2\Big)\vp_\bn+\text{l.o.t.}
\end{equation}
The coefficient of~$\vp_\bn$ in the latter formula can be simplified by noting that
\[
  \sum_{j=l(n_i)}^{l(n_i)+\#(n_i)-1}\la_j(\bn)^2=4\sum_{k=l(n_i)}^{l(n_i)+\#(n_i)-1}\Big[n_k+a(N-2k+1)\Big]^2,
\]
where we have performed the change of index $k=2l(n_i)+\#(n_i)-j-1$ and used the fact that
$n_k=n_i$ for $l(n_i)\le k\le l(n_i)+\#(n_i)-1$. We thus obtain
\begin{equation}
  \label{laisimp}
  \sum_i\la_i(\bn)^2=4\sum_i\Big[n_i+a(N-2i+1)\Big]^2=:E(\bn)\,.
\end{equation}

We shall finally compute $H'$ on a basis function~$\vp_\bn$ with an arbitrary $\bn\in\ZZ^N$. To
this end, let us denote by~$W$ any permutation such that~$\vp_\bn=W\vp_{[\bn]}$. Since the
auxiliary operator $H'$ clearly commutes with $K_{ij}$ for all $i\ne j$, it also commutes with any
permutation $W$. We thus have
\[
  H'\vp_\bn=WH'\vp_{[\bn]}=E([\bn])W\vp_{[\bn]}+W(\text{l.o.t.}) =E([\bn])\vp_{\bn}+\text{l.o.t.},
\]
where we have used the fact that the partial order~$\prec$ is invariant under permutations. We
conclude that~$H'$ is indeed represented by an upper triangular matrix in the basis~$\cB$ of
$\fH'$, with eigenvalues~$E([\bn])$ with $\bn\in\ZZ^N$. Moreover, since $H'$ preserves the
subspace~$\cP_n$ for all integers $n$, $\fH'$ admits a basis of (unnormalized) eigenfunctions of
$H'$ of the form
\begin{equation}
  \label{psin}
  \psi_\bn=\vp_\bn+\sum_{\bn'\prec\bn,\,|\bn'|=|\bn|}c_{\bn'\bn}\vp_{\bn'}\,,\qquad
  \bn\in\ZZ^N\,.
\end{equation}
Note that the sum over~$\bn'$ in the previous equation is actually finite, since
$\max_i |n'_i|\le p(\bn)$ by the invariance of the subspaces $\cR_p$ under $H'$. On the other
hand, the auxiliary operator~$H'$ (as well as its counterparts $H$ and $\Hsc$ in the previous
section) is translationally invariant, and thus commutes with the total momentum operator
\[
  P=-\iu\sum_j\frac{\pd}{\pd x_j}\,.
\]
Since obviously~$P\vp_{\bn'}=2|\bn'|\vp_{\bn'}=2|\bn|\vp_{\bn'}$, the states~\eqref{psin} are also
eigenfunctions of $P$ with eigenvalue~$2|\bn|$. It follows that the set~$\{\psi_\bn:\bn\in\ZZ^N\}$
is a (non-orthonormal) basis of~$\fH'$ consisting of common eigenfunctions of~$H'$ and~$P$, with
\begin{equation}\label{EPpsi}
  H'\psi_{\bn}= E([\bn])\psi_{\bn}\,,\qquad P\psi_{\bn}=2|\bn|\psi_{\bn}\,.
\end{equation}
Note, finally that if $\psi$ is an eigenfunction of $P$ and $H'$ with respective eigenvalues $p$
and $E$, it is straightforward to check that $\e^{2\iu\ms c|\bx|}\psi$ (where~$|\bx|:=\sum_kx_k$)
is an eigenfunction of the latter operators with eigenvalues $p+2Nc$ and $E+4cp+4Nc^2$. From this
fact and eq.~\eqref{psin} it follows that
\begin{equation}
  \label{boost}
  \psi_{\bn+c\mathbf1}(\bx)=\e^{2\iu\ms c|\bx|}\psi_{\bn}(\bx)\,,\qquad
  \mathbf 1:=(1,\dots,1)\,.
\end{equation}

\subsection{Spectrum of $H$ on $\fH$}\label{app.specHscH}

As before, we start by extending the natural Hilbert space $L^2(A)\otimes\Si$ of the Hamiltonian
$H$ of the spin dynamical model~\eqref{Hspin} to $\fH'\otimes\Si$. It is then clear from its
definition that~$H$ coincides with the auxiliary operator $H'$ in the subspace~$\fH$ of
$\fH'\otimes\Si$ defined by the relations
\[
  K_{ij}=-S_{ij},\qquad K_iK_j=S_iS_j,
\]
i.e., whose elements are i) antisymmetric under simultaneous permutations of the coordinates and
spin variables, and ii) symmetric under translations by~$\pi/2$ (modulo $\pi$) of an \emph{even}
number of particles' coordinates and reversal of their corresponding spins. This subspace is
therefore
\[
  \fH=\La_a\La_0(\fH'\otimes\Si)\,,
\]
where~$\La_a$ and $\La_0$ respectively denote the antisymmetrizer with respect to coordinate and
spin permutations and the symmetrizer with respect to the operators $K_iS_iK_jS_j$. It is
important to note that the spectrum of $H$ on $\fH$ is actually the same as its spectrum on the
more natural Hilbert space~$L^2([-\pi/2,\pi/2]\cap A)\otimes\Si$. Indeed, what we are basically
doing here is enlarging the original configuration space $A$ to $A'$ by applying the Weyl group
generated by the operators $K_{ij}$, $K_iK_j$, and then restricting ourselves to wavefunctions
defined on this larger space with a certain well-defined symmetry under the action of the
corresponding operators $K_{ij}S_{ij}$, $K_iS_iK_jS_j$ (see, e.g., ref.~\cite{BFG11} for a more
technical explanation).

The symmetrizer~$\La_0$ can be expressed in terms of the projectors~$\La_0^{\pm}$ onto states
symmetric (``$+$'') or antisymmetric (``$-$'') under the action of the operators $K_iS_i$, defined
by 
\[
\La_0^\pm=\frac1{2^N}\left(\sum_{i=1}^{2^{N-1}}W_i^+\pm \sum_{i=1}^{2^{N-1}}W_i^-\right),
\]
where~$W_{i}^+$ (respectively $W_i^{-}$) denotes the product of an even (respectively odd) number
of operators~$K_iS_i$, as
\[
  \La_0=\frac1{2^{N-1}}\sum_{i=1}^{2^{N-1}}W_i^+=\La_0^++\La_0^-\,.
\]
We thus have\footnote{Note that $\La_0^\pm$ commutes with all the permutations $K_{ij}S_{ij}$, and
  hence with $\La_a$.}
\[
  \fH=\La_a\La_0^+(\fH'\otimes\Si)\oplus\La_a\La_0^-(\fH'\otimes\Si)\,.
\]
A basis of~$\fH$ consists therefore of the states
\begin{equation}\label{Phi}
  \Phi^\vep_{\bn,\bbs}=\La_a\La_0^\vep(\vp_\bn(\bx)\ket\bbs)\,,
\end{equation}
where $\bn\in\ZZ^N$ and $\bbs\in\{-M,-M+1,\dots,M\}^N$ must be appropriately restricted so that
the latter states are linearly independent. For instance, it suffices that
\begin{enumerate}
\item $n_1\ge\cdots \ge n_N$
\item $\ds n_i=n_j\implies s_i>s_j$
\item $s_i\ge0$ for all $i$, and $s_i>0$ if $(-1)^{n_i}\vep=-1$\,.
\end{enumerate}
Indeed, the first two conditions simply take into account the antisymmetry
of~$\Phi_{\bn,\bbs}^\vep$ with respect to simultaneous permutations of spatial coordinates and
spins. As to the third one, note first of all that acting on $\Phi^\vep_{\bn,\bbs}$ with $K_iS_i$
(which does not change the physical state, since by construction
$K_iS_i\Phi^\vep_{\bn,\bbs}=\vep\Phi^\vep_{\bn,\bbs}$) we can reverse the sign of $s_i$ if
necessary. Moreover, if $s_i=0$ we have
\[
  K_iS_i\Phi^\vep_{\bn,\bbs}=(-1)^{n_i}\Phi^\vep_{\bn,\bbs}=\vep\Phi^\vep_{\bn,\bbs}\,,
\]
so that $(-1)^{n_i}\vep=1$. By the same token, the states
\[
  \Psi^\vep_{\bn,\bbs}=\La_a\La_0^\vep(\psi_\bn(\bx)\ket\bbs)\,,
\]
where the quantum numbers $\bn$ and $\bbs$ satisfy conditions~1--3 above, are also a basis
of~$\fH$. Since, again by construction,
\[
  H'\La_a\La_0^\vep=H\La_a\La_0^\vep\,,
\]
and
\[
  [H',K_{ij}]=[H',K_i]=0\en\implies\en [H',\La_a]=[H',\La_0^\vep]=0\,,
\]
we have
\begin{align*}
  H\Psi^\vep_{\bn,\bbs}
  &=H\La_a\La_0^\vep(\psi_\bn\ket\bbs)
    =H'\La_a\La_0^\vep(\psi_\bn\ket\bbs)=\La_a\La_0^\vep\big((H'\psi_\bn)\ket\bbs\big)\\
  &=\La_a\La_0^\vep\big(E(\bn)\psi_\bn\ket\bbs\big)=E(\bn)\Psi^\vep_{\bn,\bbs}\,.          
\end{align*}
Similarly,
\[
  P\Psi^\vep_{\bn,\bbs}=\La_a\La_0^\vep\big((P\psi_\bn)\ket\bbs\big)
  =2|\bn|\Psi^\vep_{\bn,\bbs}\,.
\]
Thus the states~$\Psi_{\bn,\bbs}^\vep$, with $\vep=\pm$ and~$\bn$, $\bbs$ satisfying the previous
three conditions, are a basis of~$\fH$ of common eigenfunctions of~$H$ and~$P$ with energy
$E(\bn)$ and momentum $2|\bn|$.

\section{Evaluation of the sum~\protect\eqref{sum}}\label{app.sum}
In this appendix we shall compute the sum~\eqref{sum} used in section~\ref{sec.PF} for the
calculation of the partition function of the chain~\eqref{Hchain}.

To begin with, let us define
\[
  \pi_j:=\nu_j-\frac{|\bn|}N=\nu_j-\frac1N\sum_{j=1}^{r-1}k_j\nu_j\,,\qquad 1\le j\le r\,,
\]
(with~$\nu_r=0$, cf.~eq.~\eqref{bnnu}), and note that
\begin{equation}\label{sum1}
  \sum_ip_i(N+1-2i)=\sum_{j=1}^r\pi_j\sum_{i=L_{j-1}+1}^{L_j}(N+1-2i)
  =\sum_{j=1}^r\pi_j\ell_j(N-2L_j+\ell_j)\,,
\end{equation}
where for convenience we have set $L_0:=0$, $L_r:=N$. Introducing the new variables
\[
  \tnu_j=\pi_j-\pi_{j+1}=\nu_j-\nu_{j+1}\in\NN\,,\qquad 1\le i\le r-1,
\]
and setting~$\tnu_r=\pi_r$ we have
\[
  \pi_j=\sum_{i=j}^r\tnu_i\,,
\]
and hence
\begin{equation}
  \sum_{j=1}^r\pi_j\ell_j(N-2L_j+\ell_j)=\sum_{i=1}^r\tnu_i\sum_{j=1}^i\ell_j(N-2L_j+\ell_j)\,,
  \label{sum2}
\end{equation}
where the inner sum can be easily evaluated:
\begin{align}
  \sum_{j=1}^i\ell_j(N-2L_j+\ell_j)
  &=NL_i+\sum_{j=1}^i\big[(l_j-L_j)^2-L_j^2\big]
    =NL_i+\sum_{j=1}^i\big(L_{j-1}^2-L_j^2\big)\notag\\
  &=NL_i-L_i^2=\cE(L_i)\,.
    \label{sum3}
\end{align}
Combining eqs.~\eqref{sum1}--\eqref{sum3}, and taking into account that~$\cE(L_r)=\cE(N)=0$, we
immediately obtain eq.~\eqref{sum}.

\section{Standard deviation of the spectrum of the chain \protect\eqref{Hchain}}
\label{app.sigma}

In this appendix we shall compute in closed form the standard deviation of the spectrum of the
chain~\eqref{Hchain} when the operators $S_k$ are taken as spin flip operators. To begin with,
note that
\[
  \langle\cH^2\rangle-\langle\cH\rangle^2=\langle\oH^2\rangle-\langle\oH\rangle^2\,,
\]
where
\[
  \oH=\frac14\sum_{i<j}\bigg(\frac{S_{ij}}{\sin^2\th_{ij}}
  +\frac{\tS_{ij}}{\cos^2\th_{ij}}\bigg)=: \sum_{i\ne j}(h_{ij}S_{ij}+\wth_{ij}\tS_{ij})\,,\qquad
  \th_{ij}:=\th_i-\th_j\,.
\]
The average of~$\oH$ is easily computed using eq.~\eqref{trS}, with the result
\[
  \langle\oH\rangle=\frac1{m}\,\sum_{i\ne j}(h_{ij}+\wth_{ij})\,.
\]
On the other hand, from the identities~\cite{EFGR05}
\begin{align*}
  \tr\big(S_{ij}S_{kl}\big)&=\tr\big(\tS_{ij}\tS_{kl}\big)=m^{N-2+2(\de_{ik}\de_{jl}+\de_{il}\de_{jk})}\,,
  \\
  \tr\big(S_{ij}\tS_{kl}\big)&=
                               \begin{cases}
                                 m^{N-2}\,,& m\ \text{odd}\\
                                 m^{N-2}(1-\de_{ik}\de_{jl})(1-\de_{il}\de_{jk})\,& m\
                                 \text{even}\,,
                               \end{cases}
\end{align*}
after a long but straightforward calculation we obtain
\begin{equation}\label{sitH}
  \big\langle\oH^2\big\rangle-\big\langle\oH\big\rangle^2=
  2\bigg(1-\frac1{m^2}\bigg)\sum_{i\ne j}\big(h_{ij}^2+\wth_{ij}^2\big)
  -4\big(1-\pi(m)\big)\sum_{i\ne j}h_{ij}\wth_{ij}\,,
\end{equation}
where $\pi(m)$ is the parity of $m$. The last sum is easily evaluated using eq.~\eqref{muspec}:
\begin{equation}\label{hth}
  \sum_{i\ne j}h_{ij}\wth_{ij}=\frac1{64}\,\sum_{i\ne j}\sin^{-2}\th_{ij}\cos^{-2}\th_{ij}
  =\frac1{16}\,\sum_{i\ne j}\sin^{-2}(2\th_{ij})=\frac{1}{48}\,N(N^2-1)\,.
\end{equation}
As to the first one, from the summation formula
\[
  \sum_{i\ne j}\sin^{-4}(2\th_{ij})=\frac1{45}\,N(N^2-1)(N^2+11)
\]
in ref.~\cite{CP78} and the previous equation it readily follows that
\begin{align}
  \sum_{i\ne j}\big(h_{ij}^2+\wth_{ij}^2\big)
  &=\frac1{64}\sum_{i\ne j}
    \frac{\cos^4\th_{ij}+\sin^4\th_{ij}}{\sin^4\th_{ij}\cos^4\th_{ij}} =\frac1{64}\sum_{i\ne
    j}\frac{1-2\sin^2\th_{ij}\cos^2\th_{ij}}{\sin^4\th_{ij}\cos^4\th_{ij}}\notag\\
  &=\frac14\,\sum_{i\ne j}\sin^{-4}(2\th_{ij})-\frac18\,\sum_{i\ne j}\sin^{-2}(2\th_{ij})
    =\frac{N}{360}\,(N^2-1)(2N^2+7)\,.
    \label{hthsq}
\end{align}
Substituting eqs.~\eqref{hth} and~\eqref{hthsq} into eq.~\eqref{sitH} we easily arrive at
eq.~\eqref{si2} in section~\ref{sec.specdeg}.

\acknowledgments This work was partially supported by Spain's MINECO grant~FIS2015-63966-P and
Mi\-nis\-te\-rio de Ciencia, Innovaci\'on y Universidades grant PGC2018-094898-B-I00, as well as
Universidad Complutense de Madrid's grant~G/6400100/3000. We would also like to thank the
anonymous referee for several suggestions that have contributed to improve the paper's
presentation and scope.

% \bibliographystyle{JHEP}
% \bibliography{cmprefs}

\providecommand{\href}[2]{#2}\begingroup\raggedright\endgroup

\end{document}